\documentclass[ pre, preprint,   superscriptaddress]{revtex4-1} 

\usepackage[colorinlistoftodos,prependcaption,textsize=tiny]{todonotes}

\usepackage{graphicx}
\usepackage{hyperref}
\usepackage{color}
\usepackage{amsmath} 
\usepackage{mathtools} 
\usepackage{ctable}
\usepackage{tabularx}

\newcommand{\Eq}[1]{Eq.~(\ref{#1})}
\newcommand{\Fig}[1]{Fig.~\ref{#1}}

\newcommand{\kT}{k_\mathrm{B}T}

\def\onedot{$\mathsurround0pt\ldotp$}
\def\cddot{
  \mathbin{\vcenter{\baselineskip.67ex
    \hbox{\onedot}\hbox{\onedot}}%
  }}%

\begin{document}
\title{Theoretical rheo-physics of silk: Intermolecular associations reduce the critical specific work  for flow-induced crystallisation   }  
\date{\today}
\author{Charley Schaefer}
\email{charley.schaefer@york.ac.uk}
\affiliation{Department of Physics, University of York, Heslington, York, YO10 5DD, UK}
\author{Tom C. B. McLeish}
\affiliation{Department of Physics, University of York, Heslington, York, YO10 5DD, UK}



{
}

\begin{abstract}
{  
  {Silk is a semi-dilute solution of randomly coiled associating polypeptide chains that crystallise following the stretch-induced disruption, in the strong extensional flow of extrusion, of the solvation shell around their amino acids.
  We propose that natural silk spinning exploits both the exponentially-broad stretch-distribution generated by associating polymers in extensional flow and the criterion of a critical concentration of sufficiently-stretched chains to nucleate flow-induced crystallisation.
   To investigate the specific-energy input needed to reach this criterion in start-up flow, we have coupled a model for the coarse-grained Brownian dynamics of the chain to the stochastic, strain-dependent binding and unbinding of their associations.
  Our simulations indicate that the associations hamper chain alignment in the initial slow flow, but, on the other hand, facilitate chain stretching at low specific work at later, high rates.
  We identify a minimum in the critical specific work at a strain rate just above the stretch transition (i.e, where the mean stretch diverges), which we  explain in terms of analytical solutions of a two-state master equation.
  We further discuss how the silkworm appears to exploit the chemical tunability of the associations to optimise chain alignment and  stretching in different locations along the spinning duct: this delicate mechanism also highlights the potential biomimetic industrial benefits of chemically tunable processing of synthetic association polymers. }
}
\end{abstract}

\maketitle
%
%
\newpage

\section{Introduction}

The manufacturing of both natural and artificial polymer-based fibres relies on flow-induced crystallisation in non-linear rheological conditions \cite{Graham09, Troisi17, Nicholson19, Moghadam19, Read20}.
The energy input required by this process may be significantly reduced through tailored macromolecular interactions, as exemplified by natural silk \cite{Holland12a, Schaefer20, Schaefer21A, Schaefer21B}: This protein, of which the conformation closely resembles a random coil \cite{Asakura21}, self-assembles in flow in aqueous conditions under energy requirements orders of magnitude lower than its synthetic counterparts \cite{Holland12a}. 
It has been hypothesised that flow-induced stretching of the chain disrupts a solvation layer and in turn enables crystallisation to commence.
This mechanism was employed to induce crystallisation of synthetic poly-ethylene oxide by flow at similarly low energetic requirements as silk, however, at much higher molecular weight and/or strain rates  \cite{Dunderdale20}.
The low-energy mechanism for natural silk-spinning therefore remains to be identified. 
Clues may be present in the subtle electrostatically-modified rheo-physics of associating polymers \cite{Leibler91, Colby98, Hackelbusch13, ChenQ16, Tomkovic18, ZhangZ18, Hinton19, Golkaram19,Zuliki20,Liu21}.

\begin{figure}
  \centering
  \includegraphics*[width=10cm]{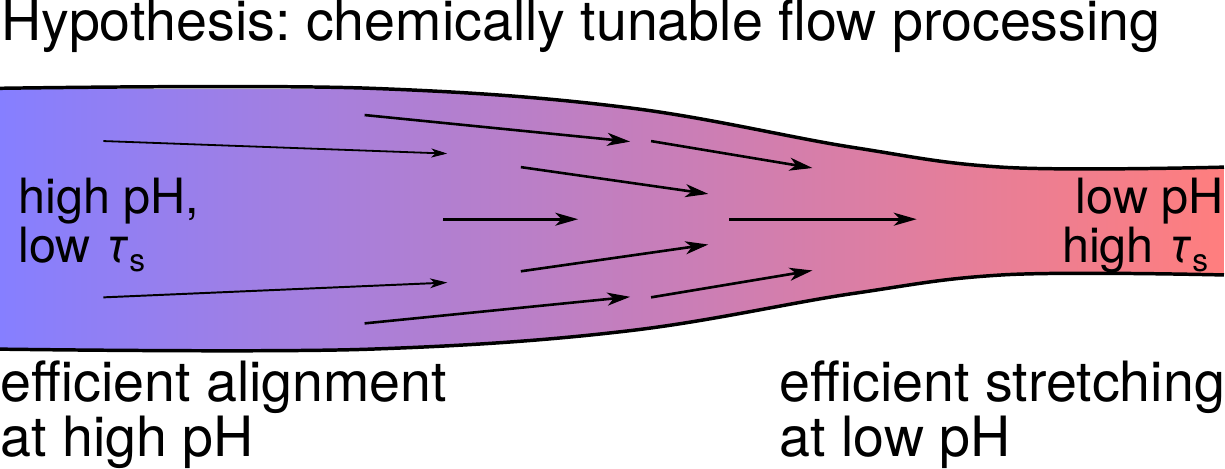}
  \caption{Schematic representation of the gland of a \emph{B. mori} silkworm, where extensional flow initially aligns at low deformation rates and subsequently stretches the intrinsically disordered silk protein at higher rates. Within the interpretation of a sticky-reptation model, we hypothesise that the experimentally observed gradients in the pH serve to control the lifetime of intermolecular cross-links locally within the process, which in turn minimises the energetic requirements to deform the network and induce fibre crystallisation.
}
\label{fig:Hypothesis}
\end{figure}

We previously found in collaboration with Laity and Holland that the silk protein exhibits intermolecular reversible cross-links \cite{Laity18, Schaefer20}.
While these associations shift the alignment-to-stretch transition to smaller strain rates by replacing the usual Rouse relaxation dynamics for `sticky Rouse' relaxation \cite{Leibler91, Colby98, Hackelbusch13, ChenQ16, Tomkovic18, ZhangZ18, Hinton19, Golkaram19,Zuliki20,Liu21}, 
this is not the full story, for the \emph{Bombyx mori} silkworm manages also to generate the opposite effect during pupation: when the silkworm starts spinning the material it actually chemically \emph{reduces} this relaxation time through the addition of potassium cations \cite{Laity18, Schaefer20}.
Intriguingly, the group of Holland discovered that the structural features of the silk fibre is significantly enhanced through a gradient in the pH along the spinning duct, suggesting an exquisitely controlled local  rheology \cite{Koeppel21}.
While lower pH may induce partial folding of the protein \cite{Asakura21}, it is also expected to enhance the lifetime of the intermolecular associations, or `stickers'.
As schematically indicated in \Fig{fig:Hypothesis}, we hypothesise {first} that the initial reduction of the viscosity decreases the specific work needed to align the chains in the direction of the flow field, while the subsequent increase of the sticker lifetime downstream promotes the stretching of chain segments, which in turn induces crystallisation.
Crucially, inspired by our previous finding that broad conformational distributions emerge due to the stochastic nature of binding and unbinding stickers \cite{ Schaefer21A, Schaefer21B}, we therefore {secondly} hypothesise that crystallisation may be initiated by reaching a critical concentration of highly stretched chain segments. This would require significantly less energy input than for stretching the entire population of chain segments.

To theoretically investigate this hypothesis, we seek to simulate the conformational response of associating (i.e, `sticky') polymers in various non-linear flow conditions, and calculate the specific critical work
\begin{equation}
  W(t_\mathrm{s})=\int_{0}^{t_\mathrm{s}}  \boldsymbol{\sigma} \cddot \boldsymbol{\kappa} \mathrm{d} t, \label{eq:SpecificWork}
\end{equation}
 needed to achieve both alignment (quantified using a nematic order parameter) and finite chain stretch in a critical fraction at time $t_\mathrm{s}$ (this integral is taken in the Lagrangian co-moving frame of a fluid element). In \Eq{eq:SpecificWork}, $\boldsymbol{\kappa}$ is the strain-rate tensor, and $\boldsymbol{\sigma}$ the stress response.
Because the stretch distribution is extremely broad for associating polymers in strong flow, it is required to capture how the entire distribution of conformational properties (rather than only the ensemble averages \cite{Likhtman02, Graham03})   evolves with time.
There are now several such classes of polymer rheology known, for which mean conformations are very poor estimates of the distribution. Other examples are linear polymers in shear \cite{Nafar15, Mohagheghi16A, Mohagheghi16B} and ring polymers in extension \cite{ HuangQ19, OConnor20, OConnor21,Bonato21}. 

The strain-rate tensor in \Eq{eq:SpecificWork} may in principle be predicted using molecular dynamics (MD) simulations, where the interactions between stickers are modelled using attractive potentials. At this level of computational detail, sticker dissociation may occur following attempts to escape the attractive potential through molecular vibrations \cite{OConnor21}.
These MD simulations are, however, computationally very demanding, as the dissociation events are quite rare. However, because of this rarity of events, the local equilibration of the chains enables a much simpler description of the chain dynamics in terms of the fraction of closed stickers, $p$ and their lifetime, $\tau_\mathrm{s}$ \cite{Leibler91}: In a coarse-grained picture, this sticker lifetime is an elementary rather than an emergent timescale.
This allows a description of the problem in terms of the dynamics of a single chain in a crowded environment \cite{Shivokhin17, Boudara17, Cui18, Schaefer21A, Schaefer21B}, an approach similar to the modelling of entangled polymers through slip-link and slip-spring models \cite{Hua98, Shanbhag01, Doi03, Schieber03, Likhtman05, Shivokhin17, Andreev20,Bonato21}, where the generation and destruction of entanglements are modelled as elementary processes.

While there is no unique way of formulating a coarse-grained single-chain model \cite{DealyBook}, all variants of bead-spring, slip-link and slip-spring models can be written in the general form 
\begin{equation}
  \zeta_i \frac{ \partial \mathbf{R}_i }{ \partial t} = \mathbf{F}_{\mathrm{intra},i} + \mathbf{F}_{\mathrm{thermal} ,i}+ \mathbf{F}_{\mathrm{flow},i} + \mathbf{F}_{\mathrm{network},i},
  \label{eq:BrownianDynamics}
\end{equation}
where $i$ is a chain segment at position $\mathbf{R}_i$ that is thermally equilibrated at the relevant time scales \cite{DoiEdwardsBook}.
We will refer to this chain segment as a `node' of an elastic network, which may represent a non-sticky segment of a chain (a regular `bead'), a segment with a reversible association (a `sticker'), or it may be an entangled segment (a `slip-link' or a `slip-spring').
Which of these representations is invoked manifests itself in the definition of the friction coefficient, $\zeta_i$, the (friction-dependent) thermal forces, $\mathbf{F}_{\mathrm{thermal} ,i}$, and the network forces, $ \mathbf{F}_{\mathrm{network},i}$.
For instance, in classes of models where nodes move affinely with the flow field, the network force exactly cancels the sum of the intramolecular force set by the chain conformation, $\mathbf{F}_{\mathrm{network},i}=-\mathbf{F}_{\mathrm{intra},i}-\mathbf{F}_{\mathrm{thermal} ,i}$.
This `rigid-network approximation' is tacitly invoked in the slip-link model by Hua and Schieber \cite{Schieber03} and in our recently published model for sticky-polymers in a rigid network  \cite{Schaefer21A, Schaefer21B}).
Within  Likhtman's slip-spring model, the slip-spring may diffuse within a potential energy landscape that represents the elastic compliance of the entangled network \cite{Likhtman05}. 
In the present work, we will account for the compliance experienced by the stickers in a reversible network.

{In the following, in Section~\ref{eq:SlipLinkModel} we develop the usual intramolecular, thermal and drag forces that act on single chains.
To capture how the stickers modify the intermolecular forces (i.e., the `elastic compliance' of the surrounding network) and the segmental drag, we present a non-spatially-explicit multi-chain approach.
In Section~\ref{sec:TwoStateModel}, we present a two-state master equation that generates analytical predictions of the impact of sticker opening and closing on both the steady-state and transient stretch distributions of the chains, which enables us to interpret our simulated data in Section~\ref{sec:Results}.
By first mapping the results in the linear flow regime to the analytic sticky-reptation (SR) model, in Section~\ref{sec:LinearDynamics} we discuss how the stochastic nature of sticker opening and closing and the elastic compliance affects the linear rheological data.
Then, in Section~\ref{sec:SteadyState} we  show how a broad steady-state distribution of chain conformations emerges in strongly non-linear flows of shear and extension.
By simulating the transient emergence of these distributions in start-up flow in Section~\ref{sec:Transients}, we show that the stickers initially hamper the collective alignments of the chains in mildly non-linear aligning flows, but facilitates the emergence of stretched outliers.
In Section~\ref{sec:W} we discuss how these outliers may reduce the critical specific work for flow-induced crystallisation. 
In the conclusions in Section~\ref{sec:Conclusions} we use our findings to interpret the experimental observations of silk spinning, and argue that the chemical tuning of associations is indeed a promising mechanism to control the flow-induced crystallisation of artificial materials.
}

\section{Model and Theory}

\subsection{Brownian dynamics of Sticky Polymers in Flow} \label{eq:SlipLinkModel}

\begin{figure}[ht!]
\centering
\includegraphics*[height=5.5cm, trim={0 0 0 0}]{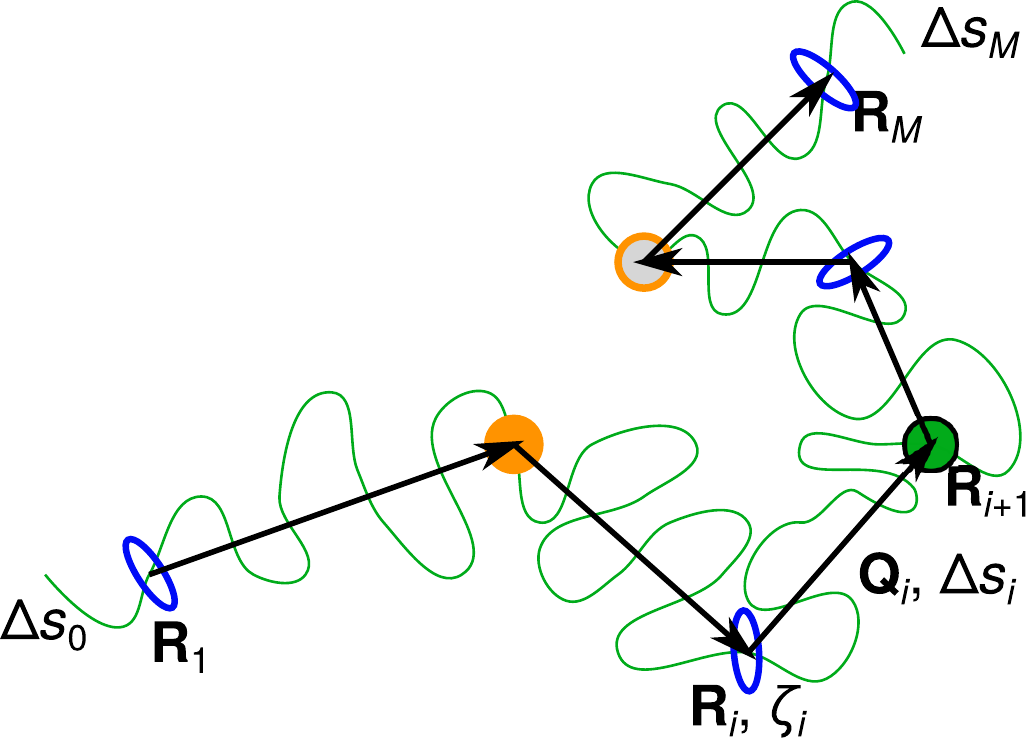}
\caption{
The theory in Section \ref{eq:SlipLinkModel} applies to sticky entangled poymers that are parameterised using the locations of $M$ nodes that may be beads (green disk), sliplinks (blue circles), closed stickers (orange disk), or open stickers (orange circles). These nodes are assigned friction $\zeta_i$ that depends on the fraction of monomers of the chain, $\Delta s_i$, that reside in each of the $M+1$ substrands, see \Eq{eq:friction}. 
In the present work, we focus on the physics of the stickers and disregard the effect of entanglements/sliplinks. 
 }
\label{fig:NodeModel}
\end{figure}

In this section we will present a coarse-grained description of associating polymers, where the dynamics of sticker opening and closing will depend on the number of open and closed stickers in a non-spatially-explicit collection of chains.  
Any linear polymer that consists of $N$ monomers may be discretised using a number  of nodes, $N_\mathrm{nodes}$, see \Fig{fig:NodeModel}.
We use the wording `node' to emphasise that the node may not just represent a traditional bead of a bead-spring model, but may also represent a sticker that can be in an open or closed state, or a slip-link or slip-spring (which, unlike traditional beads, may fluctuate in numbers).
The nodes $i$ are located at the spatial coordinate $\mathbf{R}_i$; for each chain the coordinates $\mathbf{R}_i$ are relative to the centre of mass.
The strand between neighbouring nodes $i$ and $i+1$ has an end-to-end vector $\mathbf{Q}=\mathbf{R}_{i+1}-\mathbf{R}_i$ and contains a fraction $\Delta s_i = N_{\mathrm{s},i}/(N+1)$ of all the monomers in the chain. 
(Note that the number of nodes and the number of monomers per strand fluctuate in a slip-link model, while they are usually fixed in bead-spring models.)
At this level of coarse-graining, the friction of the nodes is given by
\begin{equation}
  \zeta_i = N \zeta_0\begin{cases}
    \Delta s_{i-1} + \Delta s_{i}/2, \,\,\text{for}\,\,i=1\\
    (\Delta s_{i-1}+\Delta s_i)/2, \,\,\text{for}\,\,1<i<N_\mathrm{nodes}\\
    \Delta s_{i-1}/2 + \Delta s_{i}, \,\,\text{for}\,\,i=N_\mathrm{nodes}
  \end{cases}, \label{eq:friction}
\end{equation}
with $\zeta_0$ the monomeric friction. The assumption that the dangling chain ends are relaxed may be released by explicitly modelling the position of the chain ends and setting $\Delta s_{i}\equiv 0$ at $i=0$ and at $i=N_\mathrm{nodes}$ \cite{Andreev13}.

The equilibrium structure of the chain in quiescent conditions is determined by the end-to-end distance of the substrands, $|Q_i|=\lambda b (\Delta s_i N)^{1/2}$, where the stretch ratio $\lambda$ obeys the equilibrium distribution
\begin{equation}
  P(\lambda)=4\pi \lambda^2\left({2\pi}/3\right)^{-3/2}
                  \exp\left( - \frac{3 \lambda^2}{2}\right).
\label{eq:EquilibriumStretchDistribution}
\end{equation}
This distribution emerges as a consequence of the intramolecular and thermal forces in \Eq{eq:BrownianDynamics}.

In order to derive the intramolecular spring forces, we consider the spring force of the entire chain of $N$ monomers with a mean stretch ratio of unity
\begin{equation}
  F_{\mathrm{intra}}^{\mathrm{strand}} = \frac{3\kT}{b^2 N} k_{\mathrm{s}}(\lambda; \lambda_\mathrm{max}) (1-\lambda), 
\end{equation}
where 
\begin{equation}
  k_{\mathrm{s}}(\lambda;\lambda_\mathrm{max})=\frac{ (3\lambda_\mathrm{max}^2-{\lambda^2})/(\lambda_\mathrm{max}^2-{\lambda^2}) }{  (3\lambda_\mathrm{max}^2-1)/(\lambda_\mathrm{max}^2-1)}.
\label{eq:anharmonicity}
\end{equation}
approximately captures the anharmonicity of the spring force due to the finite extensibility of the substrand \cite{Cohen91}.  
For the substrands $i$ the terms in this equation have to be renormalised as $F_{\mathrm{intra}}\mapsto F_{\mathrm{intra},i}$, and $N \mapsto \Delta s_i N$ and $\lambda_{\mathrm{max}} \mapsto \Delta s_i^{1/2}\lambda_{\mathrm{max}} \equiv \lambda_{\mathrm{max},i}$. (a short strand has a higher harmonic spring force and a smaller maximum stretch ratio than a long chain.)
The direction of the force exerted by spring $i$ on node $i$ is ${\mathbf{Q}_{i}}/{|\mathbf{Q}_{i}|}$, while the direction of this force acted upon node $i+1$ is $-{\mathbf{Q}_{i}}/{|\mathbf{Q}_{i}|}$. Hence, the net intramolecular force exerted on node $i$ is
\begin{equation}
  \mathbf{F}_{\mathrm{intra},i} = {F}_{\mathrm{intra},i-1}^{\mathrm{strand}}\frac{\mathbf{Q}_{i-1}}{|\mathbf{Q}_{i-1}|} - {F}_{\mathrm{intra},i}^{\mathrm{strand}}\frac{\mathbf{Q}_{i}}{|\mathbf{Q}_{i}|} 
\end{equation}

The thermal force is given by the equipartition theorem  
\begin{align}
\langle \mathbf{F}_{\mathrm{thermal},i}(t)  \rangle &= \mathbf{0}; \\
\langle \mathbf{F}_{\mathrm{thermal},i,\alpha}(t)\mathbf{F}_{\mathrm{thermal},i,\beta}(t')\rangle &= 0,\, \text{for}\, \alpha\neq\beta\\
\langle \mathbf{F}_{\mathrm{thermal},i,\alpha}(t)\mathbf{F}_{\mathrm{thermal},i,\beta}(t')\rangle &= 2\kT \zeta_i \delta(i'-i)\delta(t'-t),\, \text{for}\, \alpha=\beta
\label{eq:Equipartition}
\end{align}
with $\alpha,\beta=x,y,z$ the Cartesian coordinates and $\kT$ the thermal energy.

The force acted upon the nodes by flow is, provided that our coordinate system moves with the flow field, given by 
\begin{equation}
  \mathbf{F}_{\mathrm{flow},i} \equiv \left.\zeta_i \frac{\partial \mathbf{R}_i}{\partial t}\right|_{\mathrm{flow}} = \zeta_i\boldsymbol{\kappa} \cdot \mathbf{R}_i, \label{eq:FlowForce}
\end{equation}
where $\boldsymbol{\kappa}$ is the strain-rate tensor, which in extension and shear is given by
\begin{equation}
  \boldsymbol{\kappa} = \frac{1}{2}\begin{pmatrix}
   2\dot{\varepsilon} & 0 & 0 \\
   0 & -\dot{\varepsilon} & 0 \\
    0 & 0 & -\dot{\varepsilon} 
   \end{pmatrix}, \,\, \text{and} \,\
  \boldsymbol{\kappa} =\begin{pmatrix}
   0 & \dot{\gamma} & 0 \\
   0 & 0 & 0 \\
    0 & 0 & 0 
   \end{pmatrix}, \,\,
\end{equation}
respectively.
As the coordinate system moves with the flow field, the spatial quantities of physical interest to calculate are the deformation of the individual substrands
\begin{equation}
  \left. \frac{\partial \mathbf{Q}_i}{\partial t}\right|_{\mathrm{flow}} = \boldsymbol{\kappa} \cdot \mathbf{Q}_i,
\end{equation}
using which we recursively obtain the drift of the nodes as
\begin{equation}
  \left.\frac{\partial \mathbf{R}_{i+1}}{\partial t}\right|_{\mathrm{flow}} = 
  \left.\frac{\partial \mathbf{Q}_{i  }}{\partial t}\right|_{\mathrm{flow}} + 
  \left.\frac{\partial \mathbf{R}_{i  }}{\partial t}\right|_{\mathrm{flow}}.
\end{equation}
{The value of the first entry, ${\partial \mathbf{R}_{1}}/{\partial t}$ is adjusted to fix the centre of mass of the chain (this assumes that the centre of mass moves affinely with the flow field).}


{The dynamics of the chain conformation depends on the state of the stickers through the network force, which in turn depends on the dynamics of sticker opening and closing and so, finally, on the chain conformation itself. In particular, when chain segments are highly stretched, the network forces may cause the stickers to dissociate.
To obtain these forces we simulate multiple chains and track the collection of open and closed stickers.
When sticker $i$ from chain A and sticker $j$ from chain B are closed to form a pair, the friction coefficient, the thermal force, and the network force are modified until the sticker pair opens again.
The friction coefficient of both nodes becomes $\zeta_i^\mathrm{A} + \zeta_j^\mathrm{B}$, where $\zeta_i^\mathrm{A}$ and  $\zeta_j^\mathrm{B}$ are given by \Eq{eq:friction}, and the thermal forces are given by the equipartition theorem \Eq{eq:Equipartition} as before, but with this modified friction coefficient.
The network forces are now given by
\begin{equation}
  \mathbf{F}_\mathrm{network,i}^\mathrm{A}= \mathbf{F}_\mathrm{intra,j}^\mathrm{B},\,\,\text{and by}\,\,
  \mathbf{F}_\mathrm{network,j}^\mathrm{B}= \mathbf{F}_\mathrm{intra,i}^\mathrm{A}.
\end{equation} 
Hence, the paired stickers $i$ and $j$ have an identical friction coefficient and experience the same net force $\mathbf{F}_\mathrm{intra,i}^\mathrm{A} + \mathbf{F}_\mathrm{intra,j}^\mathrm{B} + \mathbf{F}_\mathrm{thermal,i}^\mathrm{A}$ (where $\mathbf{F}_\mathrm{thermal,i}^\mathrm{A}=\mathbf{F}_\mathrm{thermal,j}^\mathrm{B}$).
Crucially to forced sticker dissociation, the net force that acts on the closed sticker pair is
\begin{equation}
  F_\mathrm{stic} = |\mathbf{F}_\mathrm{intra,i}^\mathrm{A} - \mathbf{F}_\mathrm{intra,j}^\mathrm{B}|,
\end{equation}  
which we assume, as in other cases of forces temporary unbinding, lowers the activation energy for sticker dissociation as $E_\mathrm{act} = E_\mathrm{act}^0 - \ell F_\mathrm{stic}$ with $E_\mathrm{act}^0$ the activation energy in quiescent conditions and $\ell$ the typical length scale associated with sticker dissociation \cite{Schaefer21B}.}

{
We remark that the (apparent) activation energy obtained from experiments using the Arrhenius-type equation\cite{ZhangZ18} $\tau_\mathrm{s}=\nu \exp(E_\mathrm{act}/\kT)$, for the sticker lifetime with $\nu$ an attempt frequency, may be much larger than this activation energy for dissociation. This is  due to  fast sticker recombination processes \cite{Rubinstein98B,Schaefer20} or due to the mixing of various mechanisms of sticker opening and closing, such as  bondswapping \cite{Schaefer21B}.
For now, we assume a well-defined pairwise association-mechanism, for which the rate of sticker opening is $k_\mathrm{open}=\tau_\mathrm{s}^{-1}$ and the rate of sticker closing is $k_\mathrm{close}=\tau_\mathrm{s}^{-1}(1-p)/p$ with $p$ the equilibrium fraction of closed stickers \cite{Leibler91,Schaefer21B}.
In our non-equilibrium simulations, the actual number of open stickers fluctuates and $k_\mathrm{close}$ is rescaled accordingly.
We implement the opening and closing of stickers using a kinetic Monte Carlo (kMC; also known as a Discrete Event Simulation) scheme, where after a time interval $\Delta t$ a sticker is opened or closed with a probability $(1-\exp[-k_\mathrm{open}\Delta t])$ or $(1-\exp[-k_\mathrm{close}\Delta t])$, respectively.
In the Appendix we discuss the algorithm that couples a kMC scheme for sticker opening and closing to the Brownian dynamics of the chain conformation.
This coupling is independent of the assumed mechanism of pairwise association-dissociation, and may in principle be used to also describe pairwise bondswapping and the formation of larger sticker aggregates if appropriate kMC algorithms are applied \cite{Lukkien98, HeermannBook10, JansenBook12}.
}

\subsection{Approximate theory in transient extensional flow: Two-state model}
\label{sec:TwoStateModel}

The dynamics of sticky polymers is complicated by the fact that a polymer with $Z_\mathrm{s}$ stickers can be in $2^{Z_\mathrm{s}}$ different states, as each individual sticker can be either open or closed.
An instructive simple case is a chain with  $Z_\mathrm{s}=2$, as the chain is either completely free to relax when either of the stickers is open (state 1), or can only be extended by flow when both stickers are closed (state 0).
We have previously shown that this simple `sticky dumbbell' or `two-state' model possesses the double advantage of capturing the essential qualitative physics of the more complex cases and offering some analytical tractability if the chain stretch remains below the finite extensibility limit.
In particular, this theory predicts steady-state power-law stretching distributions in extensional flow below the stretching transition of chains with various numbers of stickers \cite{Schaefer21A},
\begin{equation}
  P(\lambda) \propto \lambda^{\nu},\, \text{with} \, \nu<1,\,\text{and for} \,\lambda\gg 1. \label{eq:PowerLaw}
\end{equation}
Before embarking on simulations with the full multi-sticker and finitely-extensible chain model described in the last section, therefore, we will anticipate the phenomenology with the simple two-state model. 
In particular, we extend our previous analysis to describe the dynamics by which the two-state stretch distribution evolves in start-up flow.

The starting point is to consider a chain in two states where the stickers are either  closed (state 0)  or open (state 1).
The opening rate is $\kappa_{-}$ and the closing rate is $\kappa_{+}$.
The time development of the probability distribution of the stretch ratio is described by \cite{Schaefer21A}
\begin{alignat}{2}
  \frac{\partial P_0}{\partial t} &= -\frac{\partial}{\partial \lambda}\left[ \dot{\varepsilon}\lambda P_0 \right] &- \kappa_{-}P_0 + \kappa_{+}P_1,\\
  \frac{\partial P_1}{\partial t} &= -\frac{\partial}{\partial \lambda}\left[ \left(\dot{\varepsilon}\lambda +\frac{1-\lambda}{\tau_\mathrm{R}}\right)P_1 \right] &+ \kappa_{-}P_0 - \kappa_{+}P_1.
\end{alignat}
To  approximate this equation analytically, we reduce it to a system of first-order ordinary differential equations by  first introducing the variable $y\equiv \ln \lambda$, which using the chain rule  $(\lambda {\partial  P_i}/{\partial \lambda}) = {\partial P_i}/{\partial (\ln \lambda)} \equiv {\partial P_i}/{\partial y}$ results in 
\begin{alignat}{3}
  \frac{\partial P_0}{\partial t} &= - \dot{\varepsilon} \frac{\partial P_0}{\partial y} &- \left(\dot{\varepsilon} + \kappa_{-}\right)P_0 &+\kappa_{+}P_1, \label{eq:TwoStateEqA} \\
  \frac{\partial P_1}{\partial t} &= - \left(\dot{\varepsilon}+\mathrm{e}^{-y}-\tau_\mathrm{R}^{-1}\right) \frac{\partial P_0}{\partial y} 
&+ \kappa_{-}P_0 
&- \left(\kappa_{+}+\dot{\varepsilon}+\mathrm{e}^{-y}-\tau_\mathrm{R}^{-1}\right)P_1. \label{eq:TwoStateEqB}
\end{alignat}
The non-linear contributions can then be omitted by considering the limit of large stretches, i.e., we approximate $\mathrm{e}^{-y}\approx 0$, which is equivalent to $\lambda \gg 1$.

In steady state, the left-hand side of the equation is zero, and the solution is given by\cite{Schaefer21A}
\begin{align}
  P_0^{\mathrm{eq}} &= c \lambda^{\nu},\\
  P_1^{\mathrm{eq}} &= \frac{\kappa_{+}}{\kappa_{-}}\frac{\dot{\varepsilon}}{(\dot{\varepsilon}-\tau_\mathrm{R}^{-1})} P_0^{\mathrm{eq}},
\end{align}
with $c$ a normalisation constant (its value can in principle be determined by releasing the approximation $\mathrm{e}^{-y}\approx 0$), and with the exponent of the power-law distribution given in terms of physical parameters by
\begin{equation}
  \nu = -1 + \frac{\kappa_{+}}{(\tau_\mathrm{R}^{-1}-\dot{\varepsilon})} - \frac{\kappa_{-}}{\dot{\varepsilon}}.\label{eq:StretchingExponent}
\end{equation}
(this is one of the eigenvalues of \Eq{eq:TwoStateEqA} and \Eq{eq:TwoStateEqB}; the other eigenvalue is $-1$ and is unphysical as a distribution of the form $\lambda^{-1}$ cannot be normalised.) 
{The value of this stretching exponent diverges if the bare stretch transition at $\dot{\varepsilon}\tau_\mathrm{R}=1$ is approached from small strain rates. However, because of the physics of the stickers, actual divergence already occurs at lower strain rates: At at $(1-p)\dot{\varepsilon}\tau_\mathrm{R}=1$, the exponent becomes $\nu=-1$ and the stretch distribution can no longer be normalised.
Depending on the sticker lifetime, at smaller strain rates the exponent may reach a value $\nu=-2$ if the `sticky Weissenberg number' $(1-p)\dot{\varepsilon}$ reaches unity; here, the mean stretch diverges.
While the mean stretch is finite for smaller strain rates,  the variance of the stretch diverges for $\nu\geq -3$, which happens if $(1-p)\dot{\varepsilon}$ becomes larger than $1/2$ \cite{Schaefer21A},} at which point (considerably slower than the bare stretch transition) we expect a long tail of very high stretched chains to develop in the distribution.  

{This analytic approach can be extended to predict the transient dynamics of the distribution in start-up flow.
As we will show, the late-stage dynamics in which the tail of the distribution `fills up' is independent of the initial conditions.
In those late stages, the distribution equilibrates for stretches below a certain front, $\lambda_\ast(t)$  (above which the distribution function has a value of zero) which shifts to high stretch values over time. 
The precise number of chains with a certain stretch also  depends on the width of this moving front.
We assess analytical predictions on the front position and width using the two-state model using solutions in an early- and late-stage regime, where the time scale is, respectively, much shorter and much larger than the sticker lifetime. 
While the long-time regime will slow down the progression of the front due to sticker opening, in the early-stage regime we will obtain an upper limit of the rate by which the front moves.}

In the early-stage regime,  we approximate the stretch distribution using a the Dirac-delta distribution (justified by the very wide long-time distribution), $P_i(t=0, \lambda)=c_i\delta(\lambda-\lambda_\ast(0))$ at $\lambda_\ast(0)$, from which it can be easily seen that the distributions shift to higher stretches for the closed state, $P_0(t,\lambda)=c_0\delta(\lambda - \lambda_\ast(0)\exp[\dot{\varepsilon}t])$ and retract to smaller stretches for the open state $P_1(t,\lambda)=c_1\delta(\lambda - \lambda_\ast(0)\exp[-(\tau_\mathrm{R}^{-1}-\dot{\varepsilon})t])$. 
This suggests that the `front', $\lambda_{\ast}(t)$, of any distribution with finite $P_0$, shifts exponentially in time to higher values  through $\lambda_\ast(t)=\lambda_\ast(0)\exp[\dot{\varepsilon}t]$.

{To develop an analytic approximation for the long-time limiting approach to the steady-state,  the stretch distribution is equilibrated below the stretch ratio of the front $\lambda_\ast(t_0)$ at time $t_0$, while the tail of the distribution is empty for larger stretch ratios.
Assuming that $\lambda_\ast(t_0)\gg 1$, the equilibrated part of the distribution is virtually independent of time $t$; this provides a fixed-boundary condition at $\lambda_\ast(t_0)$. This problem essentially models the dynamical response to a unit step, and lends itself to an analysis through a  Laplace transform to give a solution for the distribution at each stretch ratio $\lambda$ of the form $\exp(-s\tau(\lambda))/s$, which is the Laplace transform of a time-dependent function that becomes non-zero at the time $\tau(\lambda)$. The inverse function $\lambda(\tau)$ is then the trajectory of the `front' of the distribution.
In the Appendix, we detail the Laplace transform of Eqs.~(\ref{eq:TwoStateEqA}-\ref{eq:TwoStateEqB}) with the boundary condition in this long-time regime, which as a solution gives}
\begin{align}
  {P}_0(t,\lambda_{\ast}(t)) &=  c \left(
\frac{\lambda_\ast(t)}{\lambda_\ast(t_0)}\right)^{\nu}\Theta(\nu'\ln[\lambda_\ast(t)/\lambda_\ast(t_0)] - \dot{\varepsilon}t)\\
  {P}_1(t,\lambda_\ast(t)) &=  \frac{\kappa_{+}}{\kappa_{-}}\frac{\dot{\varepsilon}}{(\dot{\varepsilon}-\tau_\mathrm{R}^{-1})} {P}_0(t,\lambda),
\end{align}
with $\nu$ the `equilibrium stretch exponent' in \Eq{eq:StretchingExponent} and with
\begin{equation}
\nu' =\left(1 -  \frac{1}{1-\mathrm{Wi}^0} 
+ \frac{1}{1 - \mathrm{Wi}^{\mathrm{sticky}}}\right)
\end{equation}
the `dynamic stretch exponent', which controls the growth of the front of the distribution as
\begin{equation} 
\lambda_\ast(t)=\lambda_\ast(t_0)\exp\left(\frac{\dot{\varepsilon}(t-t_0)}{\nu'}\right).\label{eq:MovingFront}
\end{equation}
{Upon approaching the stretch transition $\mathrm{Wi}^{\mathrm{sticky}}=1$ where the mean stretch diverges, $\nu'\approx 0$ indicates `critical slowing down', as the  (late-stage) front of the distribution becomes immobile.
For chains with strong stickers $(1-p)\tau_\mathrm{s}\gg \tau_\mathrm{R}$ at the strain rate $\mathrm{Wi}^{\mathrm{sticky}}=1/2$ where the variance of the stretch diverges (see discussion under \Eq{eq:StretchingExponent}), we find $\nu'\approx 1/3$, which indicates that the late-stage measure of the front is shifted from the early-stage measure for the outliers by a factor $3$.
We have also checked that the moving front is narrow for small strain rates $\mathrm{Wi}^{\mathrm{sticky}} < 1/2$.
{In the Appendix, we provide more analytical analysis of the two-state model to estimate the width of the front (relative to its extent) as $\Delta_\mathrm{rel} \propto \sqrt{p \mathrm{Wi}^0\mathrm{Wi}^{\mathrm{sticky}}/(1-{\mathrm{sticky}})}$, 
where $\Delta \approx \left(\partial [P(\lambda,t)/P_\mathrm{eq}(\lambda,\infty)]/\partial \ln \lambda\right)^{-1}/\ln \lambda$.
As we show in the Appendix, typically this width is $\Delta_\mathrm{rel}\ll 1$, and the front of the distribution is narrow even close to the stretch transition.}

\section{Results} \label{sec:Results}

\subsection{Linear dynamics}\label{sec:LinearDynamics}

We have verified the physics of our model in the linear viscoelastic regime by first simulating non-sticky chains of fixed length but a varying number of beads from  $M=4$ to $64$ (the beads are regularly along the backbone of the polymer, so $\Delta s_i = 1/(M+1)$ for all $i$).
\Fig{fig:BareLVE} shows that the choice of the number of beads has a negligible influence on the time evolution of the mean-square displacement, $\mathrm{MSD}$, and is in all cases in agreement with the theoretical prediction
\begin{equation}
  \mathrm{MSD}= 6Dt,
\label{eq:MSD}
\end{equation}
where the diffusivity, $D$, is for non-sticky polymers given by the bare Rouse diffusivity
\begin{equation}
  D_\mathrm{R}= \frac{1}{3\pi^2}\frac{\langle R_\mathrm{e}\rangle ^2}{\tau_\mathrm{R}}.
  \label{eq:RouseDiffusivity}
\end{equation} 
Moreover, the inset of \Fig{fig:BareLVE}  shows that also the end-to-end-distance is distributed according to the physical equilibrium result of \Eq{eq:EquilibriumStretchDistribution}.
 
\begin{figure}[ht!]
\centering
\includegraphics*[height=5.5cm, trim={0 0 0 0}]{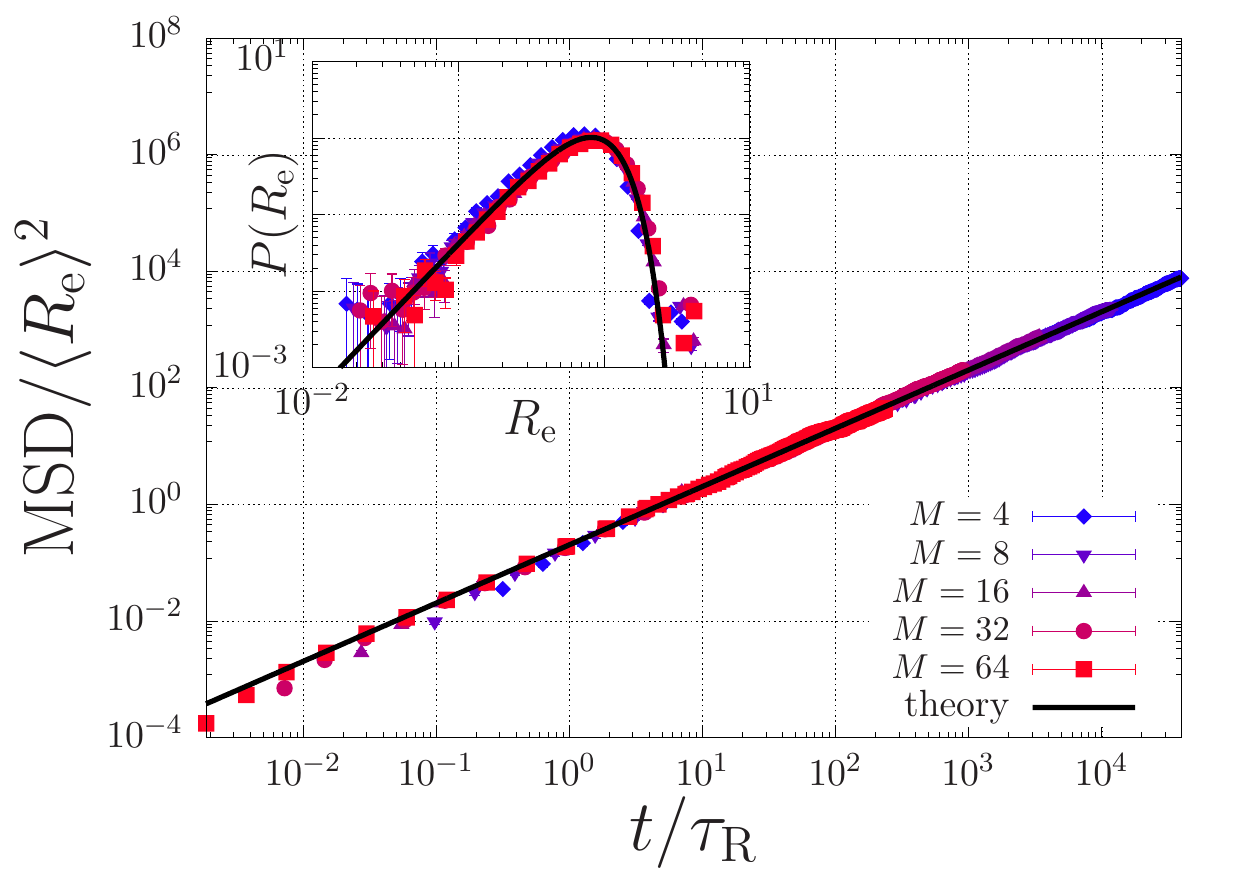}
\caption{
Mean-square displacement, $\mathrm{MSD}$, of the centre of mass of a non-sticky polymer against time (main panel)  and the time-averaged end-to-end length ($R_\mathrm{e}$) distribution (inset).
The number of real monomers per chain is fixed, while the level of coarse-graining is varied through varying the number of beads, $M$, per chain.
The symbols and solid black curves represent the simulations and the theory, respectively.
 }
\label{fig:BareLVE}
\end{figure}
 
\begin{figure}[ht!]
\centering
\hphantom{0}\hfill a)\hfill \hfill b) \hfill \hphantom{0}\\
\includegraphics*[height=5.5cm, trim={0 0 0 0}]{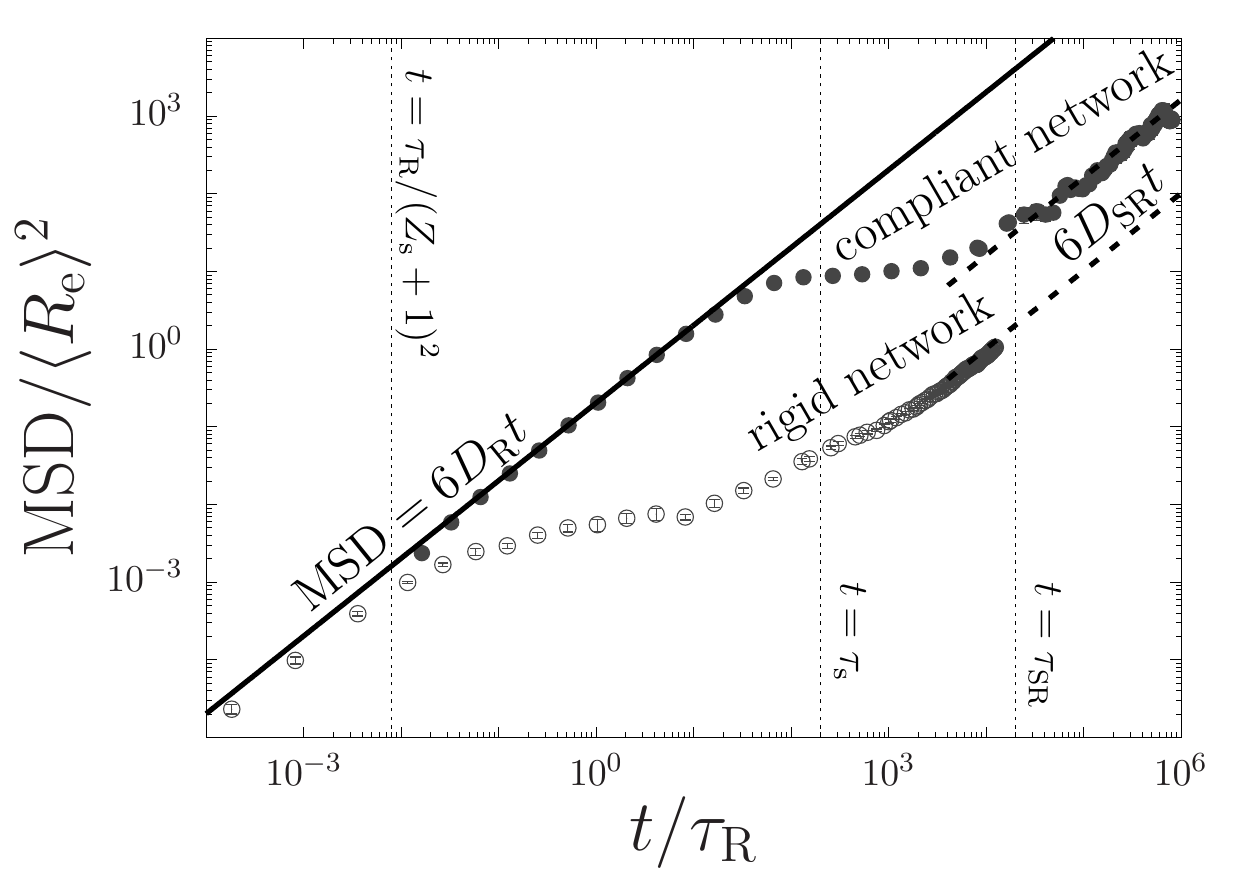}
\includegraphics*[height=5.5cm, trim={0 0 0 0}]{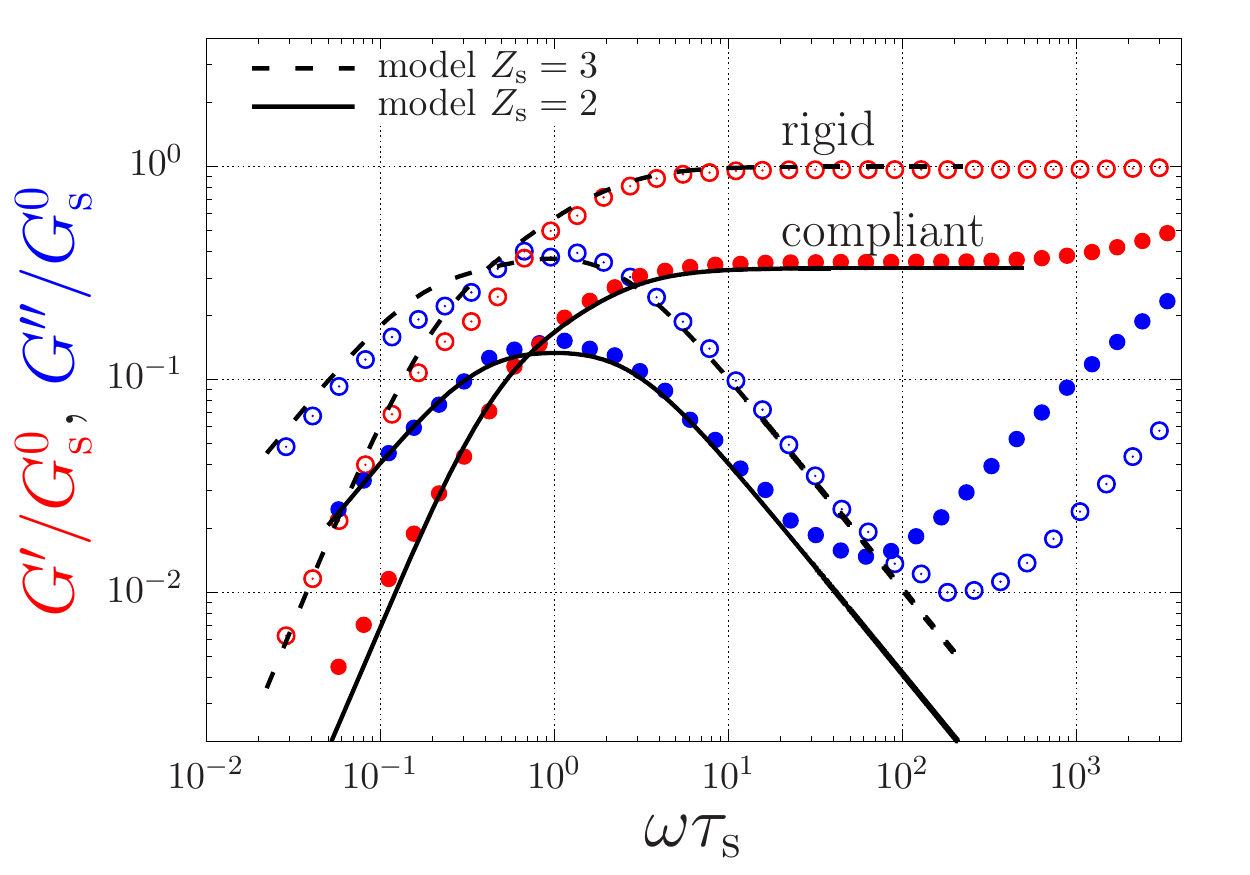}
\caption{
Linear rheology of a sticky chain with $Z_\mathrm{s}=10$, $p=0.9$, $\tau_\mathrm{s}=200\tau_\mathrm{R}$ within the rigid-network approximation (open symbols) and with this approximation released (closed symbols).
(a) Mean-square displacement $\mathrm{MSD}$ against time.
(b) Loss, $G''$,  and storage, $G''$,  moduli of the chain against the frequency, $\omega$. The moduli are scaled with respect to the plateau modulus $G_\mathrm{s}^0$ within the rigid-network approximation. The curves are the SR model of \Eq{eq:SRmodel} with   $Z_\mathrm{s}=2$ and $3$ as the \emph{apparent} number of stickers per chain.
 }
\label{fig:StickyLVE}
\end{figure}

For times shorter than the Rouse time of strands between stickers, i.e., for $t<\tau_\mathrm{R}(Z_\mathrm{s}+1)^{-2}$,  the dynamics of a sticky polymer are governed by the same Rouse diffusion as non-sticky chains, see \Fig{fig:StickyRouse}(a).
For later times than that, the motion of the polymer temporarily arrests within the rigid-network approximation (i.e., were the closed stickers are unable to diffuse, see Introduction), while it is continuous in the case the network is elastically compliant.
As the mean-square displacement increases, a larger substrand of the chain itself is engaged in the motion (requiring as in normal Rouse-chain diffusion longer Rouse wavelengths to relax), while in the case of sticky polymers  also a larger portion of the surrounding network is engaged in the motion (which increases the friction).
Consequently, the $\mathrm{MSD}$ will reach also a plateau in the compliant-network case, accounting for inter-chain interactions explicitly, but at a much higher value than within the rigid-network approximation.
As time proceeds, beyond the sticky Rouse time, $\tau_\mathrm{SR}$, sufficient numbers of stickers have opened and closed to enable longer ranged diffusive motion of the chain within the reversible network, and the MSD scales with a smaller, renormalised  diffusion constant as $\mathrm{MSD}=6D_\mathrm{SR}t$.
Within the rigid-network approximation, the sticky Rouse diffusivity, $D_\mathrm{SR}$, is given analytically by Leibler et al. in Ref. \cite{Leibler91}.
 
The consequence of the network compliance is shown in \Fig{fig:StickyRouse}(b), which shows the dynamic moduli $G'$ and $G''$ against the frequency $\omega$. 
The simulated data (symbols) where obtained from the relaxation modulus, $G(t)$,  through the multiple-tau-correlator algorithm discussed in Ref. \cite{Ramirez10}; we have obtained the dynamic moduli through the finite-element analysis discussed in Ref. \cite{Evans09}.

There appears a non-trivial mapping of equivalent parameters between this compliant-network model and analytic Rouse-chain results.
For low frequencies, the dynamic moduli obtained within the rigid-network approximation correspond to the sticky-Rouse model
\begin{equation}
  G(t) = \frac{G_\mathrm{s}}{Z_\mathrm{s}}\sum_{q=1}^{Z_\mathrm{s}}\exp\left(-\frac{2q^2 t}{\tau_\mathrm{s}Z_\mathrm{s}^2}\right),
\label{eq:SRmodel}
\end{equation}
albeit with $Z_\mathrm{s}=3$ in the analytical model, while the stochastic model has $Z_\mathrm{s}=10$. 
A similar discrepancy was previously discussed in Ref.\cite{Cui18}.
More importantly, we see an even stronger deviation if the network is compliant instead of rigid: The plateau modulus decreases and the (apparent) plateau in $G''$  narrows down.
This is in agreement with \Fig{fig:StickyRouse}(a), which suggests that the apparent (rheological) number of stickers becomes smaller than the actual number of stickers counted from in the polymer sequence.
Indeed, the number of stickers of the silk protein determined from its linear rheology was smaller by a similar factor than the number of stickers counted from the sequence of amino acids \cite{Schaefer20}.

\begin{figure}[ht!]
\centering
\hphantom{0}\hfill a)\hfill \hfill b) \hfill \hphantom{0}\\
\includegraphics*[height=5.5cm, trim={0 0 0 0}]{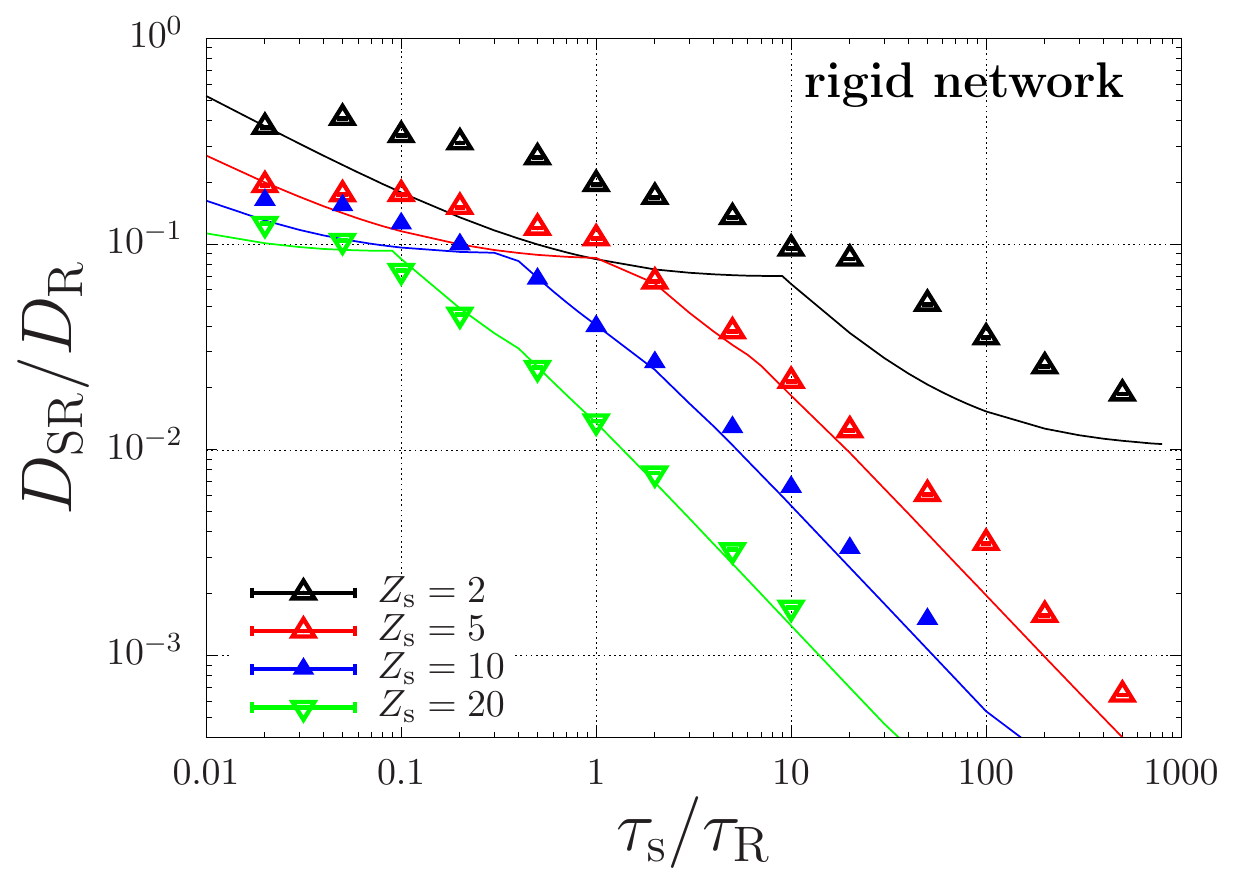}
\includegraphics*[height=5.5cm, trim={0 0 0 0}]{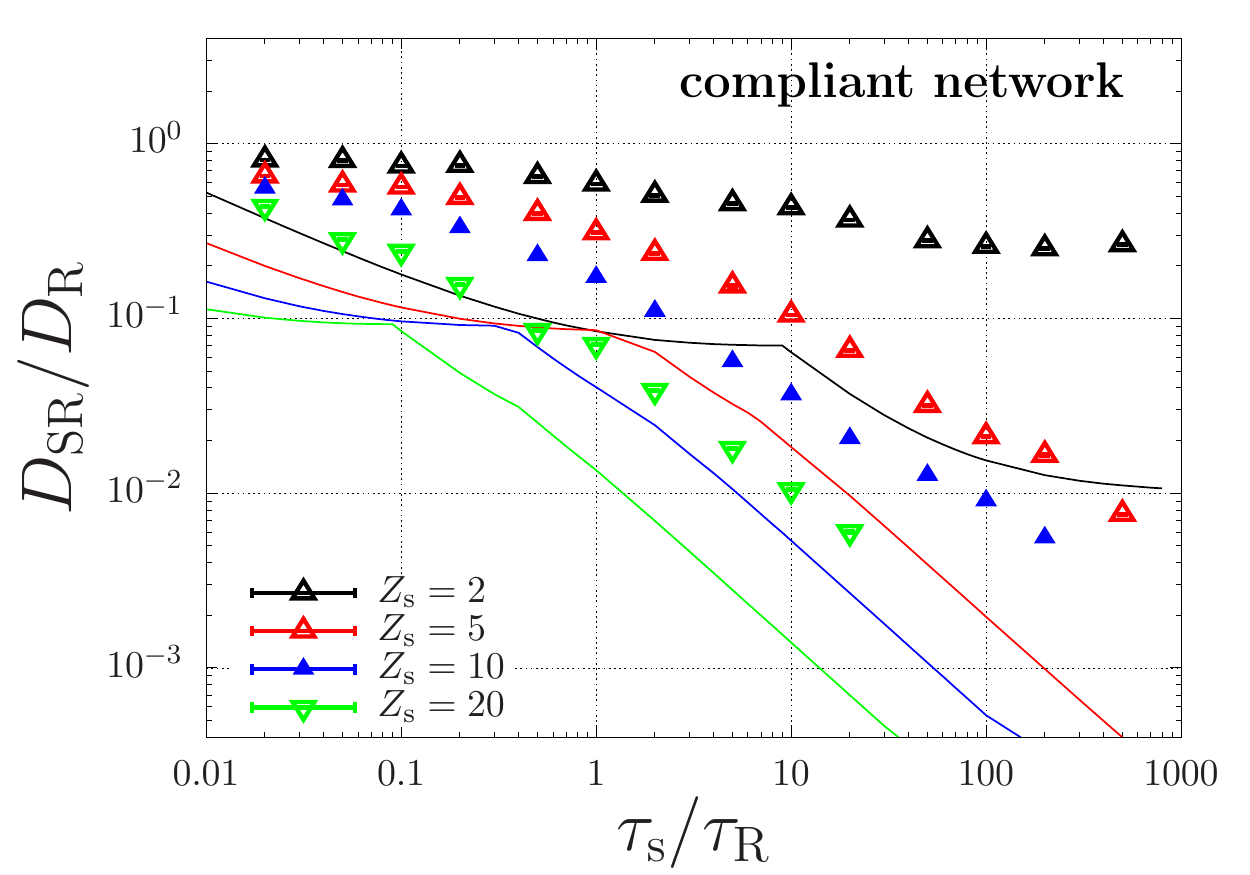}
\caption{
Sticky Rouse diffusivity, $D_\mathrm{SR}$, against the sticker lifetime, $\tau_\mathrm{s}$ for chains with $Z_\mathrm{s}=2,5,10,20$ stickers within a rigid network (a) and a compliant one (b).
The symbols are our simulation results, and the curves represents the sticky Rouse model  in Ref. \cite{Leibler91}.
The units are given in terms of the bare Rouse diffusivity $D_\mathrm{R}$ and the bare Rouse time, $\tau_\mathrm{R}$.
 }
\label{fig:StickyRouse}
\end{figure}

\Fig{fig:StickyRouse}(a) shows that the analytical model of Ref. \cite{Leibler91} describes our simulations well for chains with $5,10,20$ stickers with various sticker lifetimes, 
in particular in the regime where the sticky-Rouse diffusivity scales with the sticker lifetime as
\begin{equation}
  D_\mathrm{SR} = D_\mathrm{R}\frac{\tau_\mathrm{R}}{\tau_\mathrm{s}Z_\mathrm{s}^2}
\left(1-\frac{9}{p}+\frac{12}{p^2}\right), 
\end{equation}
see Ref.~\cite{Leibler91}.
\Fig{fig:StickyRouse}(b) shows that upon releasing the rigid-network approximation this scaling behaviour persists, but rescaled with a prefactor $\approx 4$.
While this scaling regime is reached for the chains with more than $5$ stickers (i.e., above the percolation threshold for network formation), this is not the case for the chains with $2$ stickers.
Within the rigid-network approximation, this originates from the fact that at sticker lifetimes a plateau is reached where the chains with all stickers open dominate the dynamics.
Without the rigid-network approximation, the chains cluster into linear `supramolecular' dimers, trimers, etc. through an exponentially decaying cluster-size distribution \cite{deGreef09}, which implies a distribution of diffusivities that strongly differs from that predicted by the sticky-Rouse model.
Hence, while our simulation approach accounts for the elastic compliance of the percolating network, it also captures the contributions of cluster diffusion near and below the percolation threshold for network formation.

\subsection{Non-Linear Dynamics: Steady State}\label{sec:SteadyState}

While ordinary Gaussian polymer melts and solutions of narrow molecular-weight distribution exhibit correspondingly narrow stretch distributions in steady extensional flow, in shear their conformational distributions become wide due to  dynamic stretching, tumbling and recoiling of the chains.
In this section, we investigate what these distributions look like when stickers are implemented, where we take finite extensibility of the chains and elastic compliance of the surrounding network into account.
We illustrate a striking example of how the stickers affect the steady-state chain conformations in \Fig{fig:Conformations}.
While intuitively stickers help to stretch the chains, we find that above the bare stretch transition $\dot{\varepsilon}\tau_\mathrm{R}=2$, where non-sticky chains are all stretched, the sticky chains exhibit an enormous dispersity in the chain stretch, as well as occasional hairpin conformations.
These are cause by the stochastic binding and unbinding of stickers, where the network forces may occasionally act in the opposite direction of the flow field. This mechanism rensembles the tumbling of non-sticky polymers in shear flow, but has an entirely different origin \cite{Nafar15, Mohagheghi16A, Mohagheghi16B}.

\begin{figure}[ht!]
\centering
\includegraphics*[height=6.0cm, trim={0 0 0.5cm 0}]{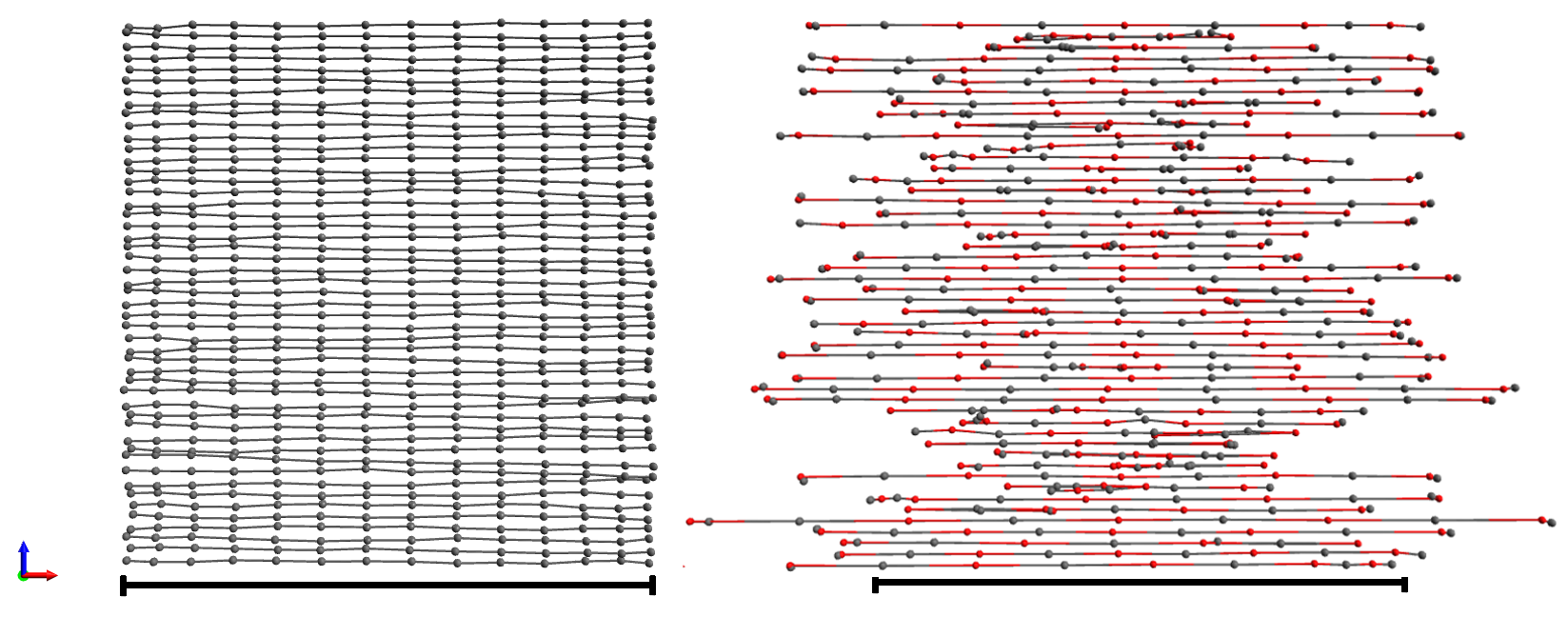}
\caption{Representation of simulated chain conformations in extensional flow for $\dot{\varepsilon}\tau_\mathrm{R}=2$ for non-sticky (a) and sticky (b) polymers. While the variations in stretch are narrow for non-sticky polymers, these variations are broad for the sticky polymers: when a sticker in a retracting chain segment binds to a neighbouring chain segment, this may disrupt the neighbouring chain. The scale bar represents approximately a length $50 R_\mathrm{e}$, which is $65\%$ of the fully extended chain. 
 }
\label{fig:Conformations}
\end{figure}


To go beyond these qualitative observations, we have calculated the stretch distributions of these chains with $Z_\mathrm{s}=0$ and $Z_\mathrm{s}=5$ at various extension and flow rates in \Fig{fig:StretchDistributions}.
In order to enable a quantitative comparison with the analytic results on the two-state model, we have also included results for a chain with $Z_\mathrm{s}=2$.

\begin{figure}[ht!]
\centering
\hphantom{0}\hfill a)\hfill \hfill b) \hfill \hphantom{0}\\
\vspace{-0.3cm}
\includegraphics*[height=6.0cm, trim={0 0 0.5cm 0}]{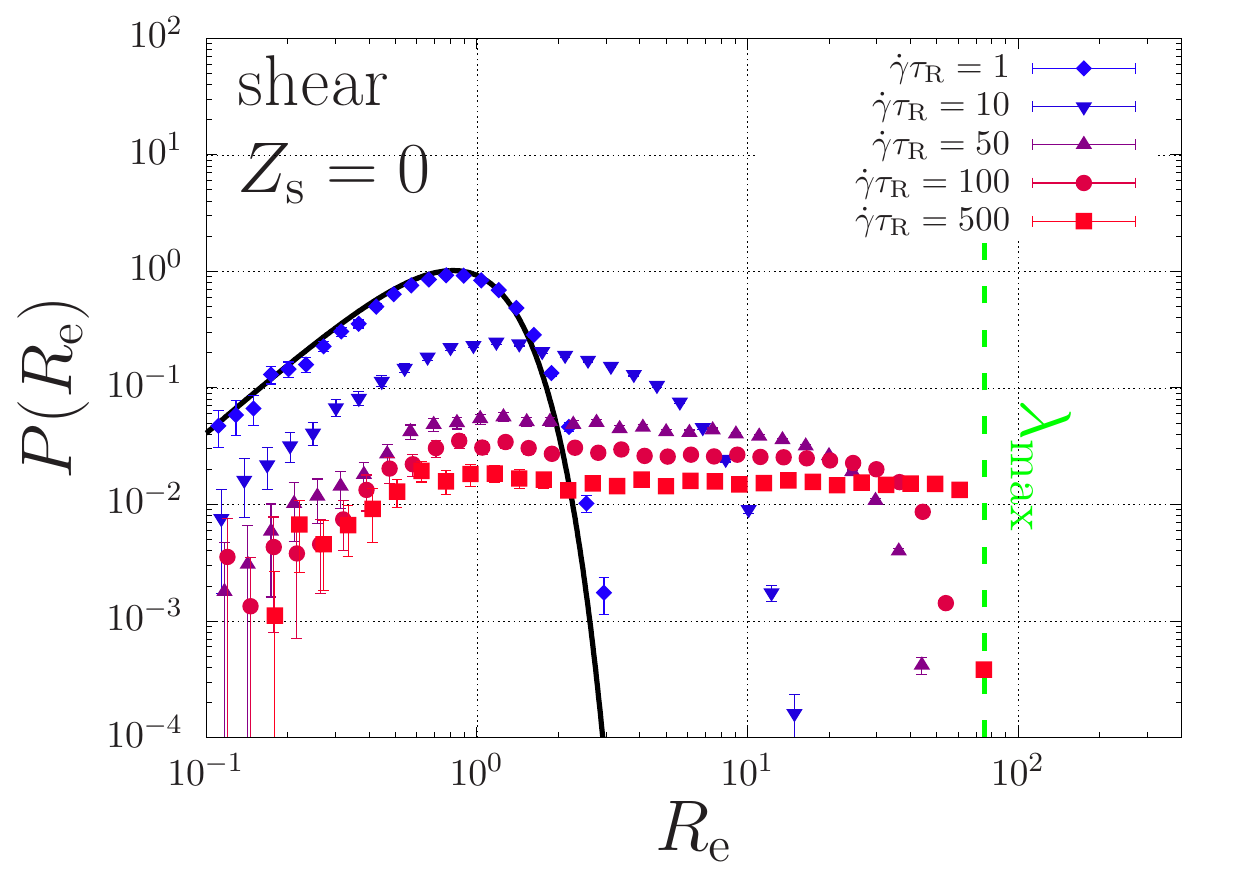}
\includegraphics*[height=6.0cm, trim={1.1cm 0 0.50cm 0}]{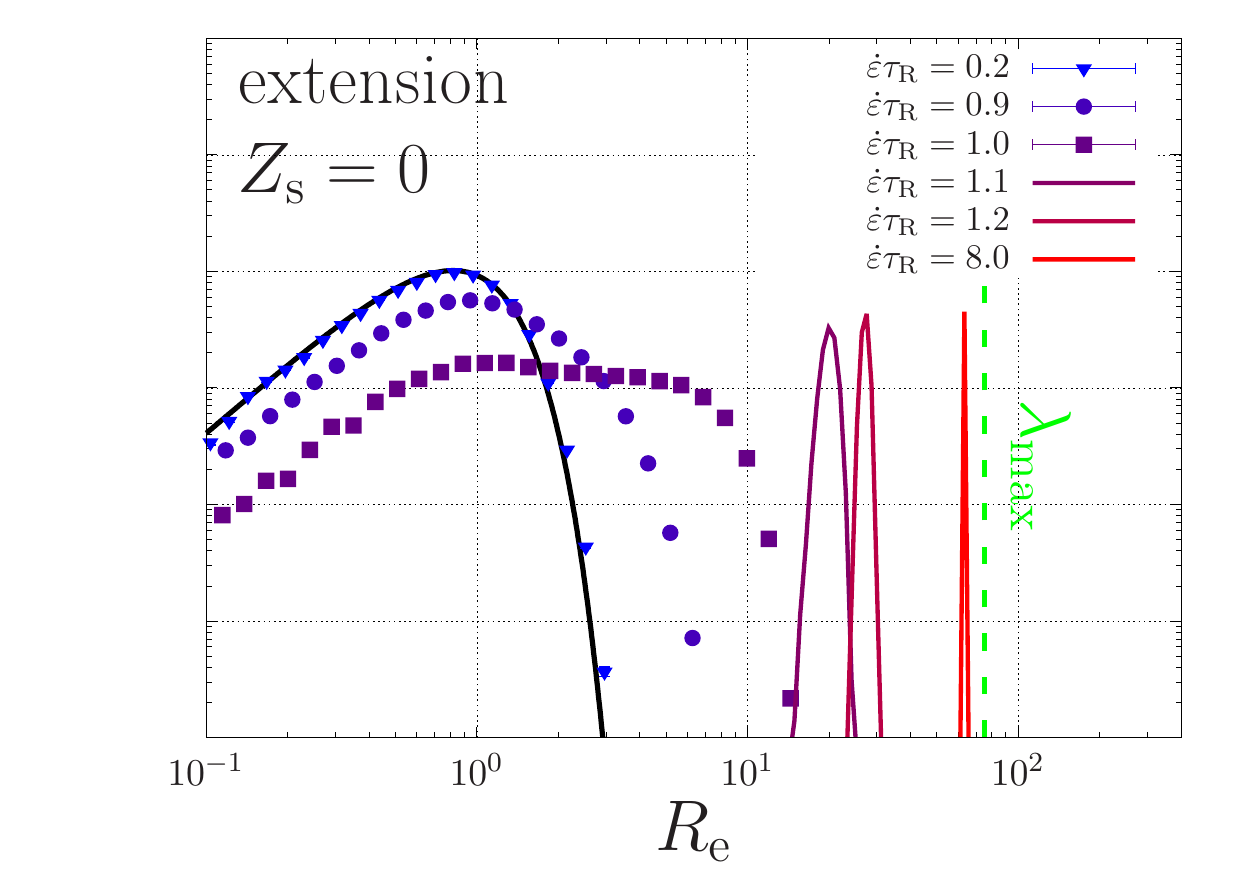}\\
\vspace{-1.5cm}
\hphantom{0}\hfill c)\hfill \hfill d) \hfill \hphantom{0}\\
\vspace{-0.3cm}
\includegraphics*[height=6.0cm, trim={0 0 0.5cm 0}]{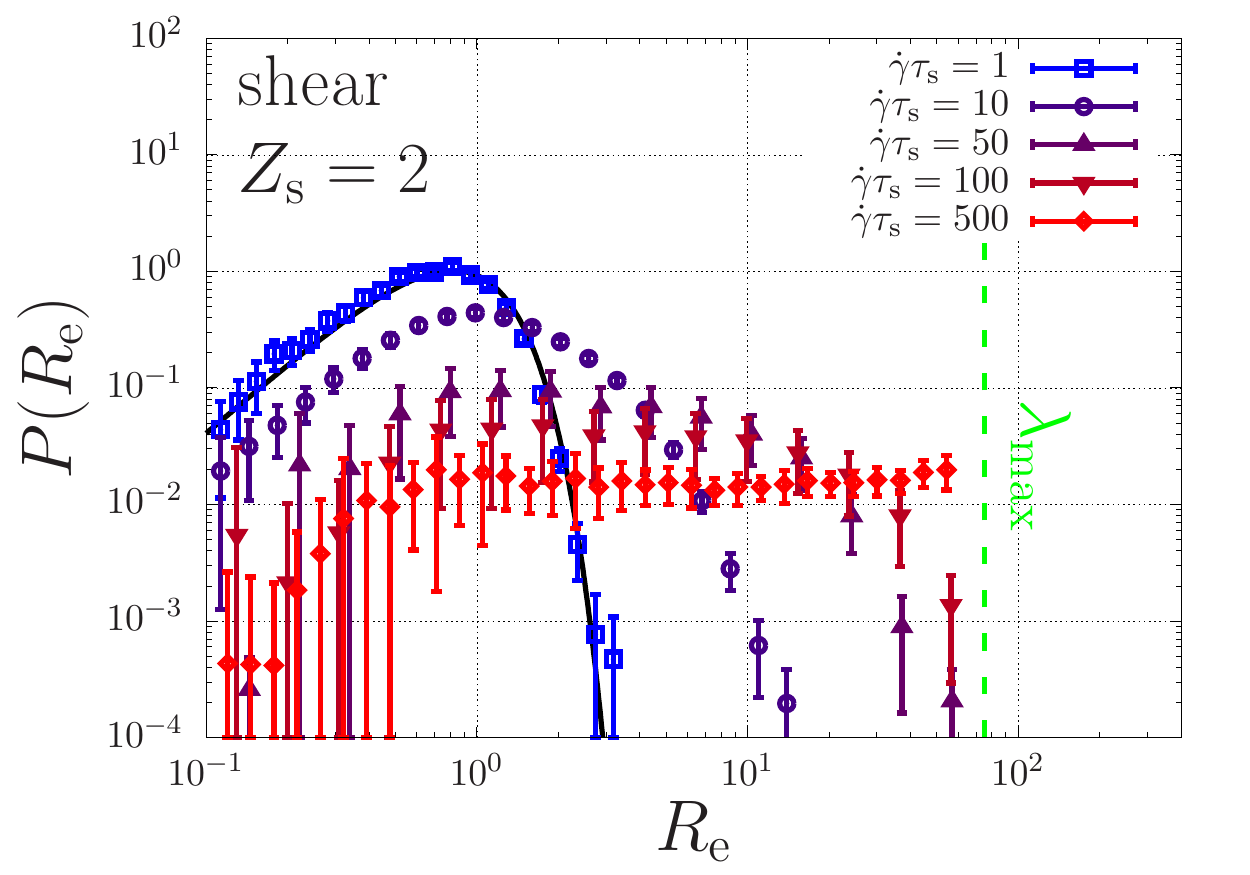}
\includegraphics*[height=6.0cm, trim={1.1cm 0 0.50cm 0}]{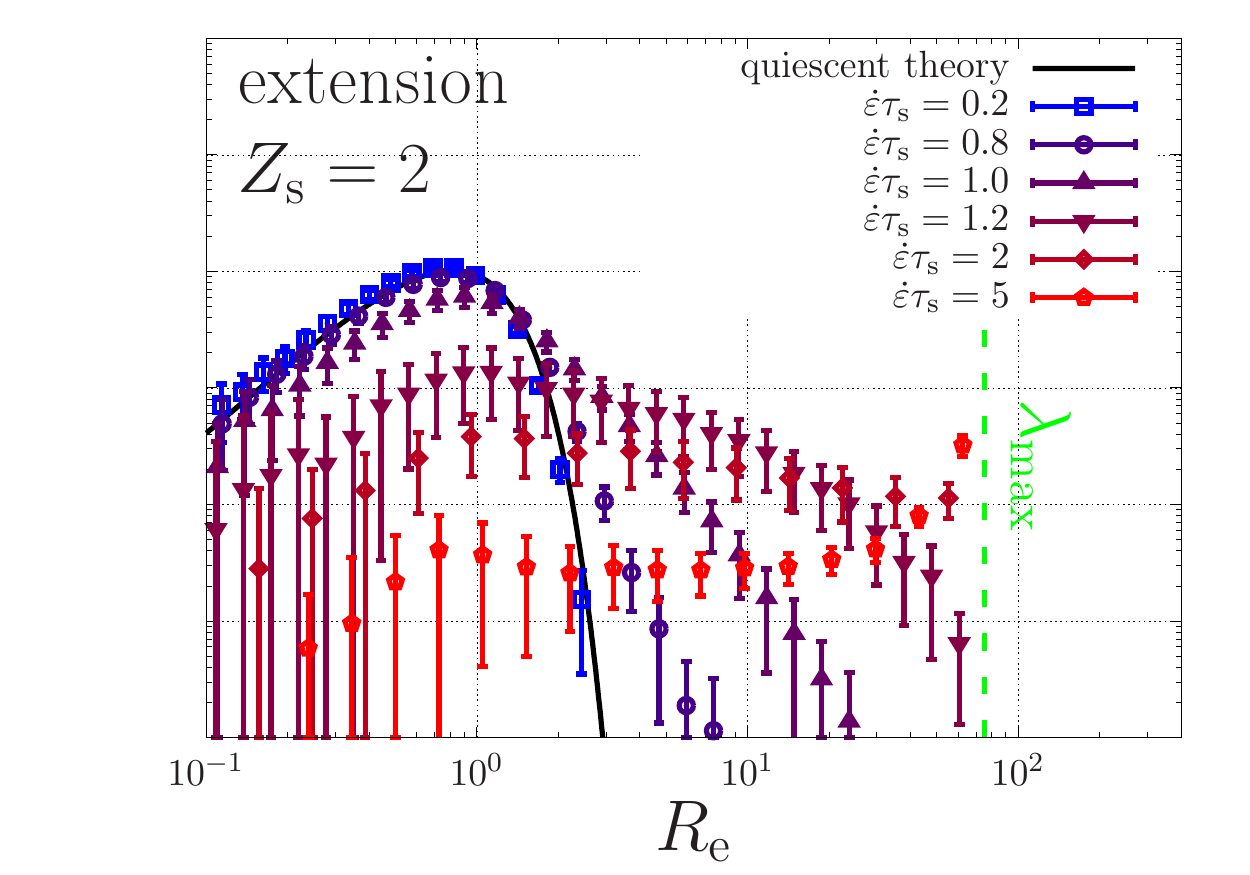}\\
\vspace{-1.5cm}
\hphantom{0}\hfill e)\hfill \hfill f) \hfill \hphantom{0}\\
\vspace{-0.3cm}
\includegraphics*[height=6.0cm, trim={0 0 0.5cm 0}]{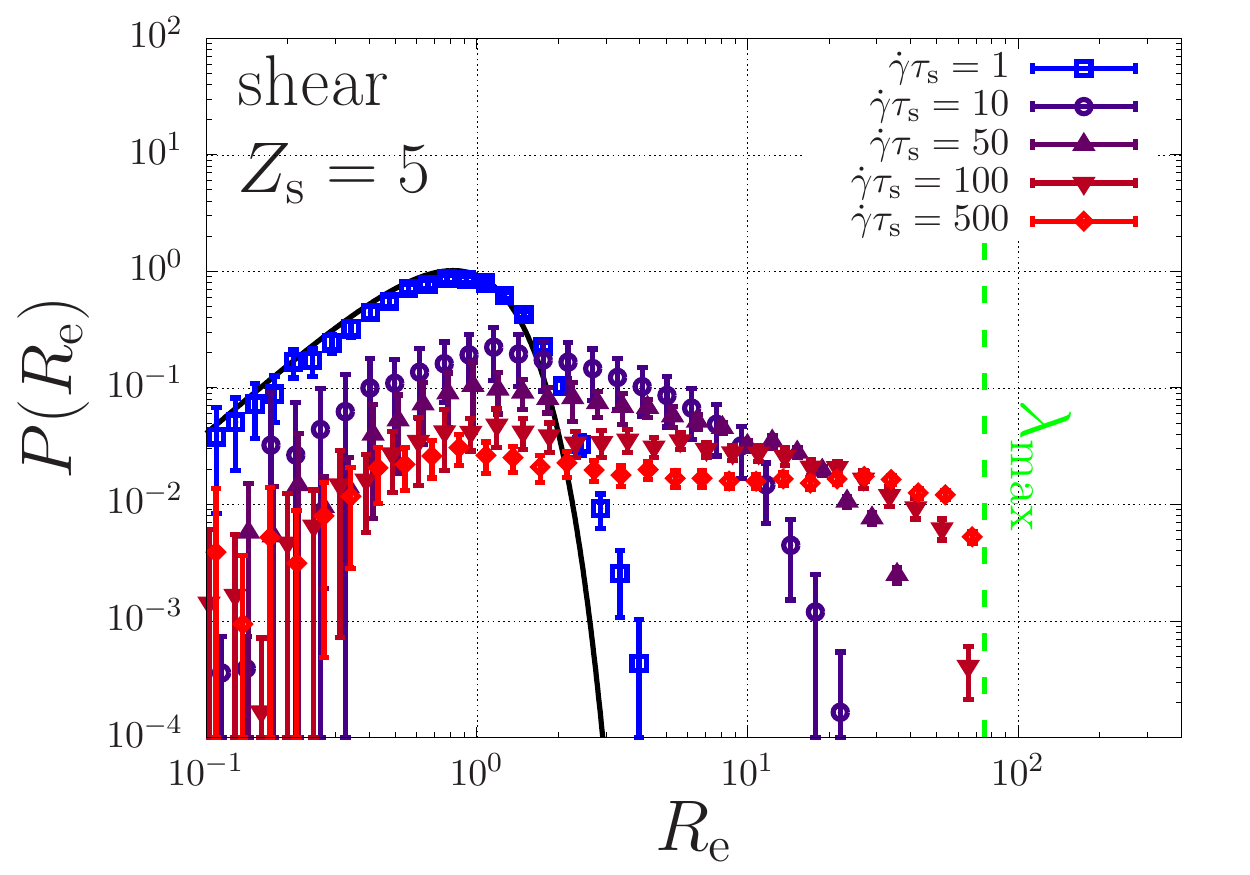}
\includegraphics*[height=6.0cm, trim={1.1cm 0 0.50cm 0}]{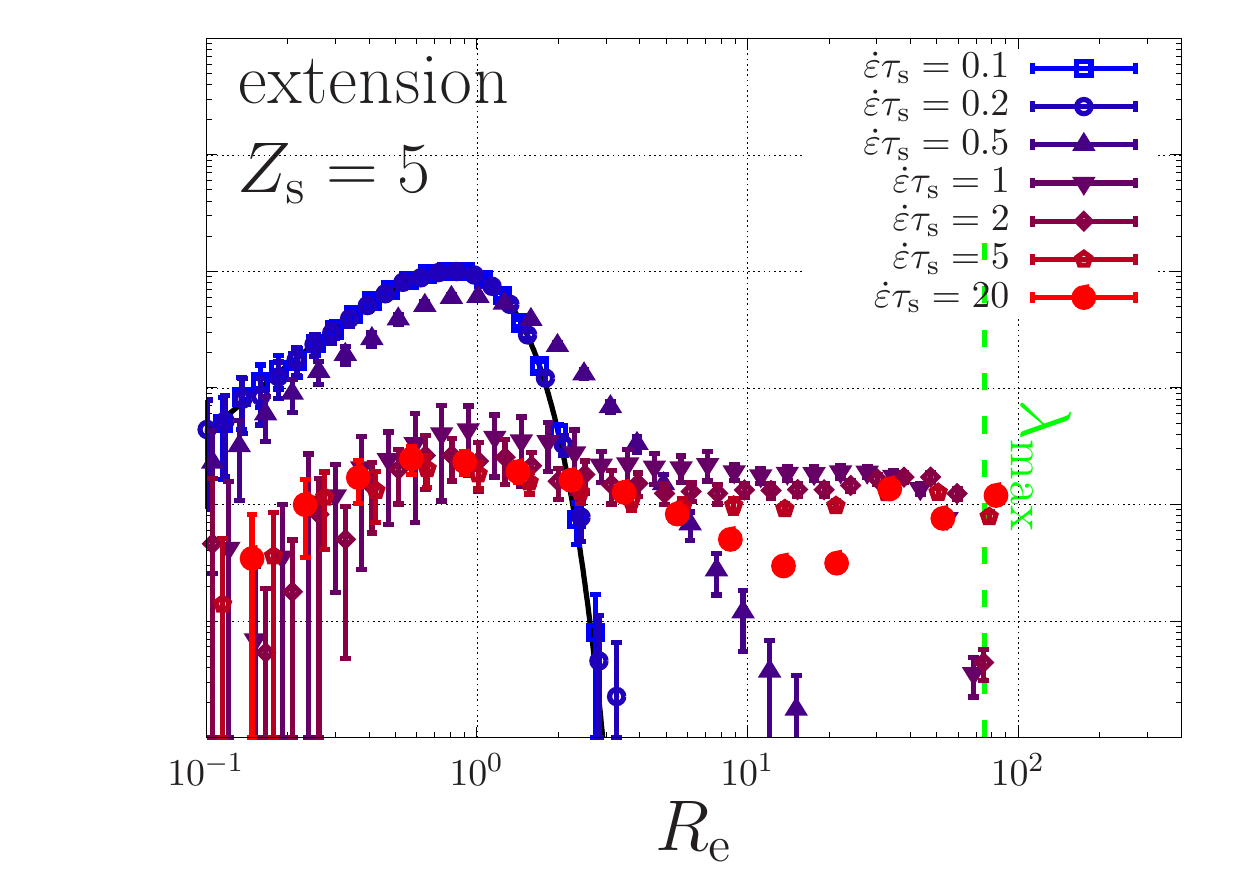}
\caption{Simulated steady-state stretch distributions for various extension (a,c,e) and shear (b,d,f) rates for a linear unentangled, non-sticky ($Z_\mathrm{s}=0$) and sticky ($Z_\mathrm{s}=2$ and $Z_\mathrm{s}=5$) polymers with a maximum stretch ratio $\lambda_\mathrm{max}=75$. The black curve represents the quiescent contour-length fluctuations, given by \Eq{eq:EquilibriumStretchDistribution}.   }
\label{fig:StretchDistributions}
\end{figure}

\Eq{eq:EquilibriumStretchDistribution} shows that in all cases the equilibrium stretch distribution for zero-flow conditions (black curve) is approached for small strain rates.
For non-sticky chains ($Z_\mathrm{s}=0$), a broad stretch distribution with a cutoff set by $\lambda_\mathrm{max}$ emerges in shear due to the dynamic stretching, tumbling and re-collapsing of the chains.
In extensional flow, the distribution broadens only within a narrow range of strain rates $0.9<\dot{\varepsilon}\tau_\mathrm{R}<1.1$ around the stretch transition.
Beyond the stretch transition, the stretch distribution is narrow and Gaussian and approaches $\lambda_\mathrm{max}$ with an increasing strain rate. This behaviour qualitatively changes upon incorporating stickers.

Polymers with $2$ or $5$ stickers form supramolecular branches below and above the percolation threshold for network formation, respectively.
Although the steady-state stretch distributions in shear are similar to those of the non-sticky chains, in  extensional flow the distributions of sticky polymers are remarkably different.
In contrast to the non-sticky polymers, the sticky polymers show broad stretch distributions in steady-state extensional flow over a broad range of flow rates.
We have observed this behaviour previously in simulations where the chains were pre-aligned in the flow-field and were we invoked the rigid-network approximation  \cite{Schaefer21B}.
Our current simulations show that this phenomenon persists when these approximations are released, but also show a dynamic coexistence of stretched chains, relaxed coils, and hairpins.

We also find that the large fluctuations in stretch below the formal stretch transition carry over \cite{Schaefer21B}.(The stretch transition is defined at the condition $\tau_\mathrm{SR}\dot{\varepsilon}=1$, with the sticky Rouse time obtained from the sticky-Rouse diffusivity of \Fig{fig:StickyRouse} as $\tau_\mathrm{SR}=\tau_\mathrm{R}D_\mathrm{SR}/D_\mathrm{R}$) 
In particular, we find that for small strain rates and large stretch ratios $\lambda$ the stretch distribution has a power-law tail (see \Eq{eq:PowerLaw}) of which the width is set by a $\dot{\varepsilon}$-dependent stretch exponent $\nu$ (see Section~\ref{sec:TwoStateModel}).
We have determined the stretch exponent from the distributions of the chains with $2$ and $5$ stickers (we discuss the numerical method in the Appendix) in extensional flow with and without the rigid-network approximation and finite extensibility, and plot these against the strain rate in \Fig{fig:StretchPower}.
As anticipated, we were able to map the stretch exponent of the chain with two stickers onto the analytical result in \Eq{eq:StretchingExponent}. To achieve that, it has to be taken into account that the open state of the chain can be achieved by opening either of the stickers; hence, $\tau_\mathrm{s}$ in \Eq{eq:StretchingExponent} is replaced by $\tau_\mathrm{s}/2$, which gives
\begin{equation}
  \nu=-1 - \frac{1}{(1-\dot{\varepsilon}\tau_\mathrm{R})}\frac{p}{(1-p)}\frac{2\tau_\mathrm{R}}{\tau_\mathrm{s}} +
\frac{2}{\dot{\varepsilon \tau_\mathrm{s}}}.  \label{eq:StickyDumbbell}
\end{equation}
We further confirmed this agreement by also simulating a chain with two stickers and $p=0.5$, and comparing it to this theory.
For chains with multiple stickers, no such analytic theory is yet available; however, we do find a qualitative agreement of the increasing power-law exponent with an increasing strain rate.

\begin{figure}[ht!]
\centering
\includegraphics*[height=6.0cm, trim={0 0 0.5cm 0}]{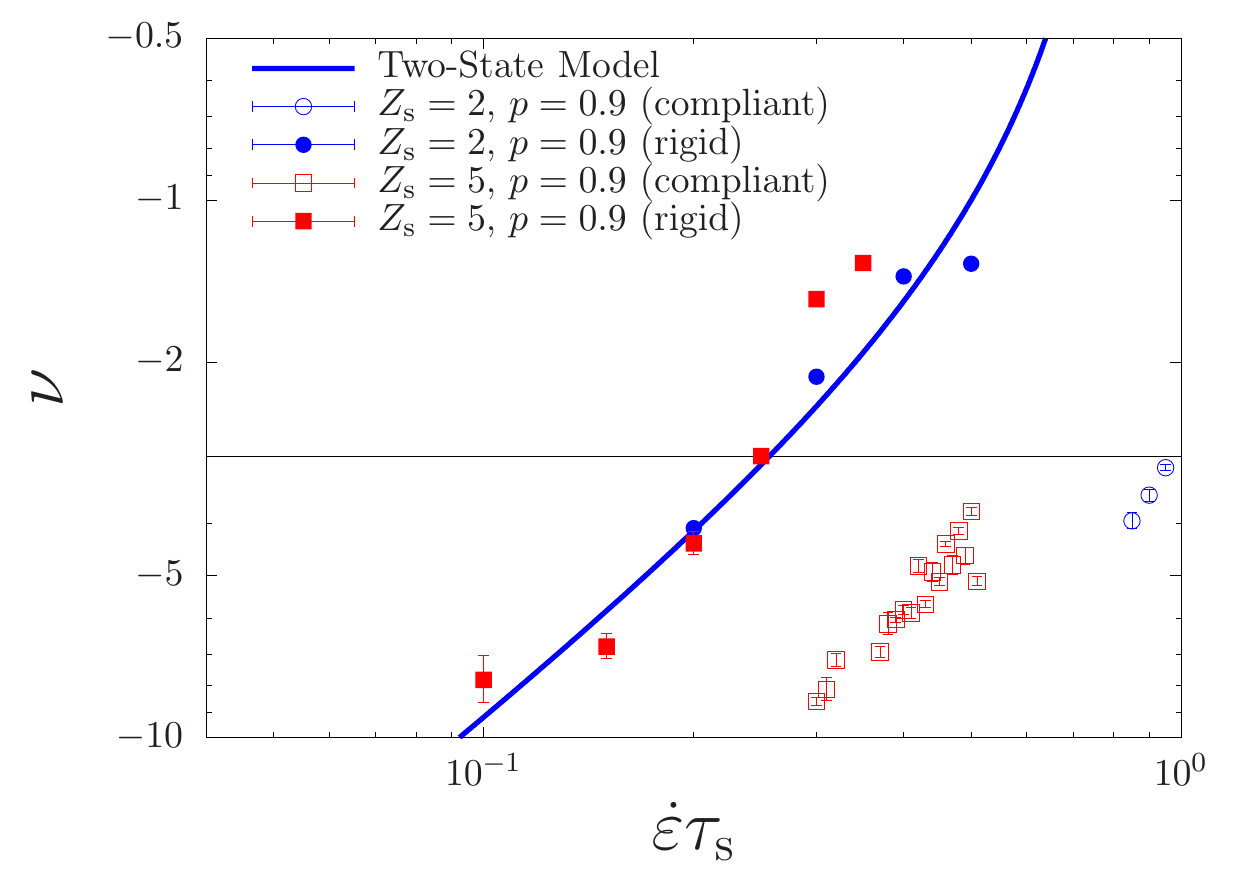}
\caption{Stretch exponent $\nu$ of the power-law tail of the stretch distribution $P\propto \lambda^{\nu}$ for simulations of polymers with $Z_\mathrm{s}=2$ (blue symbols) and $5$ stickers (red symbols) with $p=0.9$, within the rigid-network approximation (closed symbols) and using elastic compliance and finite chain extensibility (open symbols). The solid curve is given by the sticky dumbbell model in \Eq{eq:StickyDumbbell}. For $\nu>-3$ (horizontal line) the fluctuations in stretch diverge; this leads to a cutoff in the stretch distribution for chains with finite extensibility, see \Fig{fig:StretchDistributions}.  }
\label{fig:StretchPower}
\end{figure}

For the chains with $2$ and $5$ stickers and with a fraction $p=0.9$ of closed stickers, we also simulated the stretch distributions while including finite extensibility and an elastically compliant network.
Finite extensibility implies that there is a cutoff of the power-law tail, which becomes apparent with increasing (less negative) $\nu$. Since the fluctuations in $\lambda$ diverge for $\nu\geq -3$, this cutoff has a significant effect on the tail of the stretch distribution upon approaching $\nu=-3$. 
\Fig{fig:StretchPower} does confirm a broadening power-law stretch distribution for the chains in a compliant network, but shifted to higher strain rates, as expected from the faster sticky-diffusion rates from   \Fig{fig:BareLVE}.

\subsection{Non-Linear Dynamics: Transients }\label{sec:Transients}

In our pursuit to understand the flow-induced crystallisation of associating polymers such as the silk protein, we are interested in capturing the macroscopically observable stresses in start-up flow, and to interpret crystallisation rates in terms of the chain conformations that underlie these stresses.
To address these challenges, in this section we will present the time-dependent  rate-normalised transient shear stress, $\sigma_{xy}/\dot{\gamma}$, and extensional stress $(\sigma_{yy}-\sigma_{rr})/\dot{\varepsilon}$, with the stress tensor (in units of energy per molecule) given by
\begin{equation}
  \sigma_{\alpha\beta} = \frac{3\kT}{b^2 N} \sum_{i=1} \Delta s_{i-1} k_{\mathrm{s},i}\frac{Q_{\alpha,i}}{\Delta s_{i-1}} \frac{Q_{\beta,i}}{\Delta s_{i-1}}.\label{eq:StrainTensor}
\end{equation}

\begin{figure}[ht!]
\centering
\hphantom{0}\hfill a)\hfill \hfill b) \hfill \hphantom{0}\\
\includegraphics*[height=5.5cm, trim={0 0 0 0}]{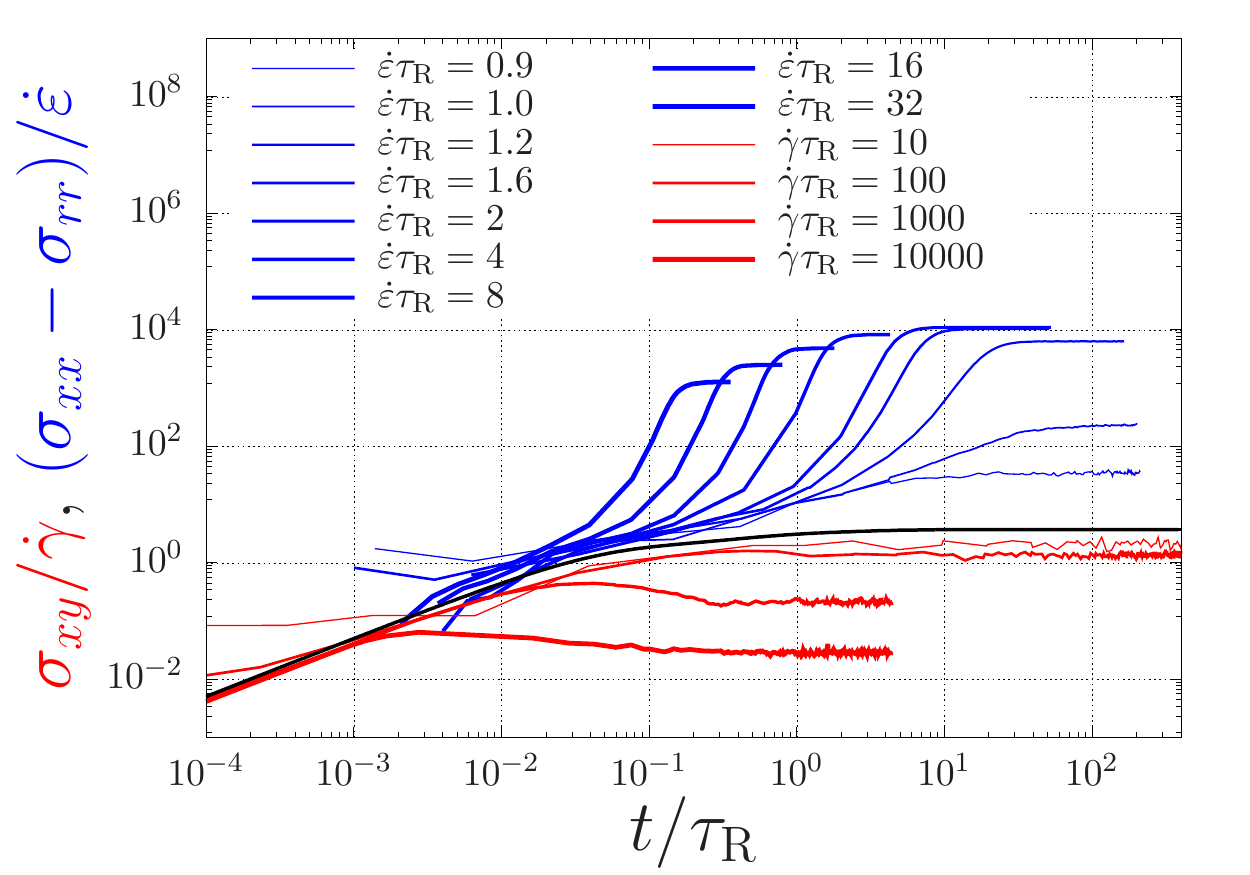}
\includegraphics*[height=5.5cm, trim={0 0 0.50cm 0}]{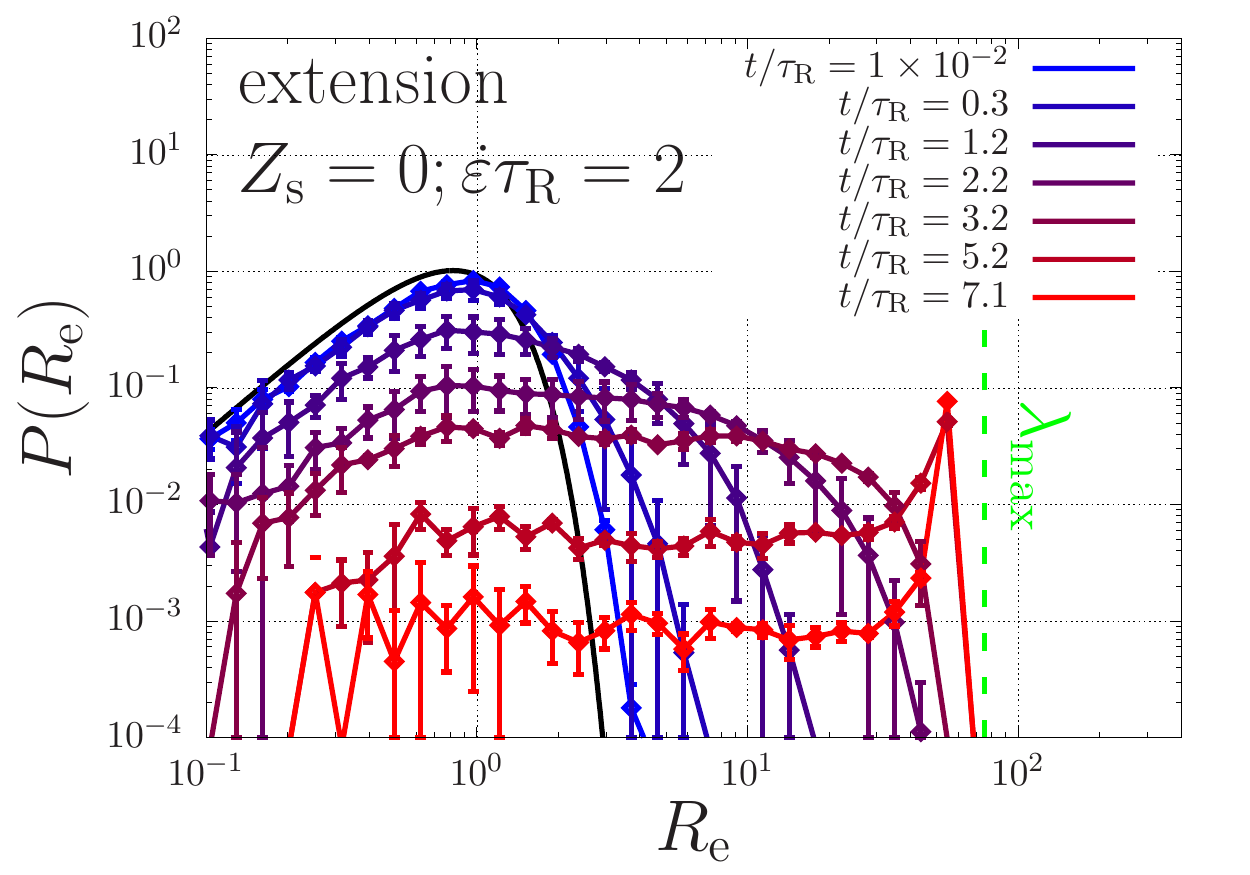}
\caption{(a) Simulated rate-normalised transient extensional and shear stresses of non-sticky polymers with a maximum stretch ratio $\lambda_\mathrm{max}=75$ (averaged over 50 polymers). (b) Transient stretch distribution in extensional flow for the chain in (a) at strain rate $\dot{\epsilon}\tau_\mathrm{R}=2$.
 }
\label{fig:BareTransient}
\end{figure}

Focussing first on the results for non-sticky chains with a finite extensibility $\lambda_\mathrm{max}=75$ in \Fig{fig:BareTransient}, we reproduce the well-known qualitative features of their stress transient\cite{DealyBook}:
For small Weissenberg numbers, $\dot{\varepsilon}\tau_\mathrm{R}<1$, $\dot{\varepsilon}\tau_\mathrm{\gamma}<1$ the polymers are able to relax, while for large strain rates there is an overshoot in shear flow, which is related to the onset of tumbling and re-collapsing of stretched chains, and in extensional flow there is a sharp increase in the extensional stress until a plateau due to the finite extensbility of the chains is reached.
Because of the thermal fluctuations and dispersity in the initial chain conformations,  \Fig{fig:BareTransient}(b) shows broadening of the stretch distribution at early times. At late times, when all chains are aligned (at the level of the beads), a sharp peak emerges at high stretches near the maximum extensibility $\lambda_\mathrm{max}$.   

This sharp peak in the stretch distribution is a fingerprint for non-sticky linear polymers in extensional flow, and will not be visible for the sticky polymers, as we we will now show for $Z_\mathrm{s}=5$. 
We have modelled the physics of the stickers using the same description as in our previous work on chains that are pre-aligned in the flow field \cite{Schaefer21B}: 
 the fraction of closed stickers is $p=0.9$, the sticker lifetime is $\tau_\mathrm{s}=10\tau_\mathrm{R}$, and the activation energy is $E_\mathrm{act}=8\kT$. Sticker dissociation is described using a dissociation length of $\ell=1$ nm; the intramolecular force to obtain the change in activation energy is obtained by assuming a chain of $N=5525$ monomers and Kuhn length $0.4$ nm). 
We plot the resulting start-up stresses and stretch distributions in \Fig{fig:TransientDistribution}.

\begin{figure}[ht!]
\centering
\hphantom{0}\hfill a)\hfill \hfill b) \hfill \hphantom{0}\\
\includegraphics*[height=5.5cm, trim={0 0 0.50cm 0}]{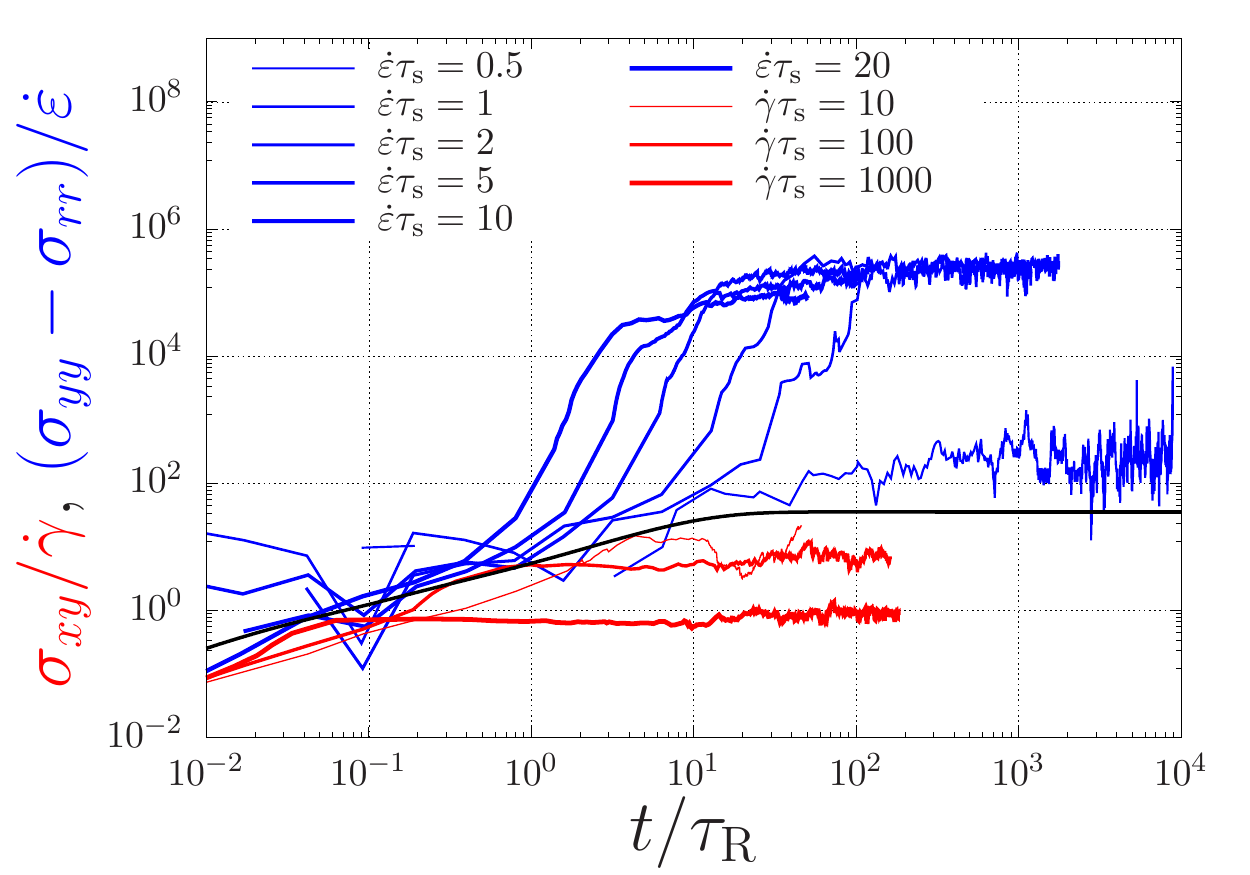}
\includegraphics*[height=5.5cm, trim={0 0 0.50cm 0}]{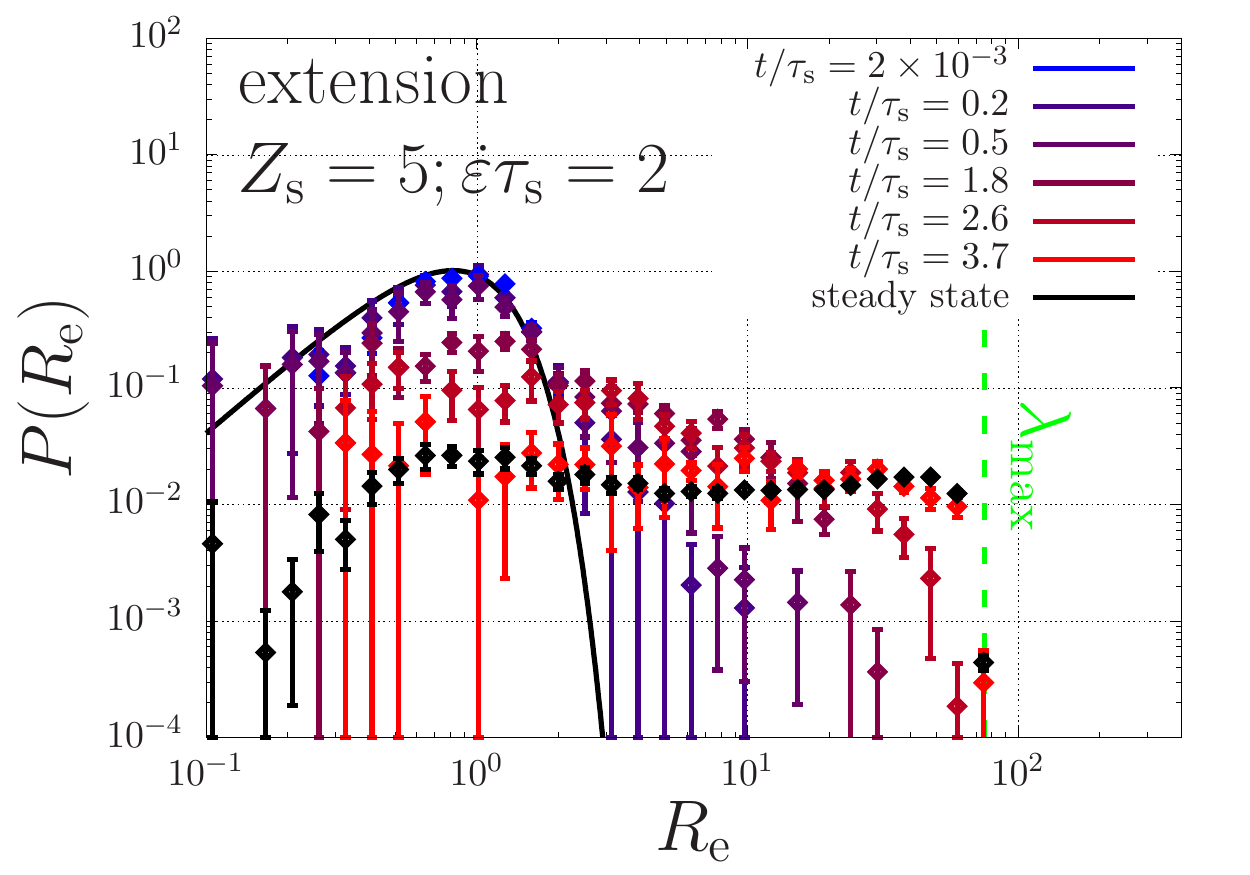}
\caption{(a) Rate-normalised transient extensional and shear stresses of sticky polymers with a maximum stretch ratio $\lambda_\mathrm{max}=75$ (averaged over 50 polymers). The sticky polymers have $Z_\mathrm{s}=5$ stickers with lifetime $\tau_\mathrm{s}=10\tau_\mathrm{R}$ and closed fraction $p=0.9$. Below the stretch transition large fluctuations are visible.
(b) Transient stretch distribution in extensional flow for the chain in (a) at strain rate $\dot{\epsilon}\tau_\mathrm{s}=2$; the error bars represent half of the standard error of the mean.}
\label{fig:TransientDistribution}
\end{figure}

Qualitatively, we find similar shear and extensional viscosities as in the non-sticky case, although there is now no distinctive overshoot in shear flow.
In extensional flow, the stresses at long time scales have shifted to higher values because of the contribution by the reversible cross-links.
Further, while non-sticky polymers show strain hardening only for $\dot{\varepsilon}\tau_\mathrm{R}>1$, the sticky ones also show strain hardening for smaller strain rates $\dot{\varepsilon}\tau_\mathrm{s}>1$. 
For strain rates smaller than that we identify large fluctuations in the transient extensional stress, as expected \cite{Schaefer21A}.
For strain rates $0.3<\dot{\varepsilon}\tau_\mathrm{s}<0.5$ these fluctuations fill up a power-law distribution whose stretch exponent is depicted in \Fig{fig:StretchPower}. For higher rates, the finite extensibility causes a truncation of this power law tail.

The dynamics by which the stretch distributions evolve in extensional flow above the stretch transition ($\dot{\varepsilon}\tau_\mathrm{s}=2$) is shown in \Fig{fig:TransientDistribution}(b). 
At early times, the stretch distribution is equilibrated according to \Eq{eq:EquilibriumStretchDistribution}.
As time proceeds, a the distribution broadens exponentially with time as $\ln \lambda \propto \dot{\varepsilon} t$ until the steady state is reached after a time $\dot{\varepsilon}\tau \propto \ln \lambda_\mathrm{max}$. 
This is in qualitative agreement with the predictions of the two-state model that we derived in \Eq{eq:MovingFront} of Section~\ref{sec:TwoStateModel}.

\subsection{Critical specific work}\label{sec:W}
{Now that we have captured how stickers lead to broad stretch distributions, we will investigate how these distributions affect the critical work for flow-induced crystallisation (FIC).
The usual predictor for FIC is the \emph{Kuhn segment nematic order parameter},  $P_{2,\mathrm{K}}\in[0,1]$.
If $P_{2,\mathrm{K}}\rightarrow 1$ (see e.g. \cite{Nicholson19}), virtually all chains are aligned at the level of the Kuhn segments, i.e., they are completely extended/stretched in the direction of the flow field.
We argue that FIC is achieved above a critical fraction of chain segments, $P_\mathrm{s}$, that are stretched above some critical stretch $\lambda_{\mathrm{s},i}\equiv L_\mathrm{s}\lambda_{\mathrm{max},i}$, with $\lambda_{\mathrm{max}_i}=\lambda_{\mathrm{max}}\sqrt{\Delta_{i}}$ the maximum stretch ratio of a segment (note that in this definition we assume ${\Delta_{i}}$ to be constant for all chain segments $i$, see Section~\ref{eq:SlipLinkModel}), and $L_{\mathrm{s}}\in[0,1]$ a critical stretch parameter, which may be viewed as a proxy for chain stretch at the Kuhn length. 
This $2$-parameter family of critical criteria is formally captured using}
\begin{equation}
  \int_{L_{\mathrm{s}}\lambda_{\mathrm{max},i}}^{\lambda_{\mathrm{max},i}} P(\lambda,t_\mathrm{s}) \mathrm{d}\lambda
 \ge P_\mathrm{s}. \label{eq:Criterion}
\end{equation}

{In our simulations, we have monitored the maximum stretch of a single chain segment in an ensemble of $50$ chains that each have $10$ chain segments (i.e., $\Delta s_i=1/10$ and $P_\mathrm{s}=1/500$; we averaged the results over $5$ simulations with different random number seeds).
At this level of coarse graining, the highest resolution of nematic chain alignment is captured using  the nematic order parameter $P_{2,\mathrm{s}}\in[0,1]$, which is the largest eigenvalue of the nematic order tensor $\mathbf{P}_{{2,\textit{\textbf{s}}}}=(3\langle \mathbf{uu}\rangle-1)/2$, where $\mathbf{u}$ is the unit vector tangential to the backbone of the chain (note that this is an overestimate of the Kuhn segment nematic order, i.e., $P_{2,\mathrm{s}}>P_{2,\mathrm{K}}$).
In \Fig{fig:SpecificWork}, we have calculated the critical specific work, $W$, as given in \Eq{eq:SpecificWork}, needed to achieve values of $P_{2,_\mathrm{s}}$ and $L_\mathrm{s}$ in the range from $0$ to $1$ for non-sticky ($Z_\mathrm{s}=0$) and sticky ($Z_\mathrm{s}=5$) chains for various shear and extensional rates. }

\begin{figure}[ht!]
\centering
\includegraphics*[height=5.8cm, trim={0 0 0.5cm 0}]{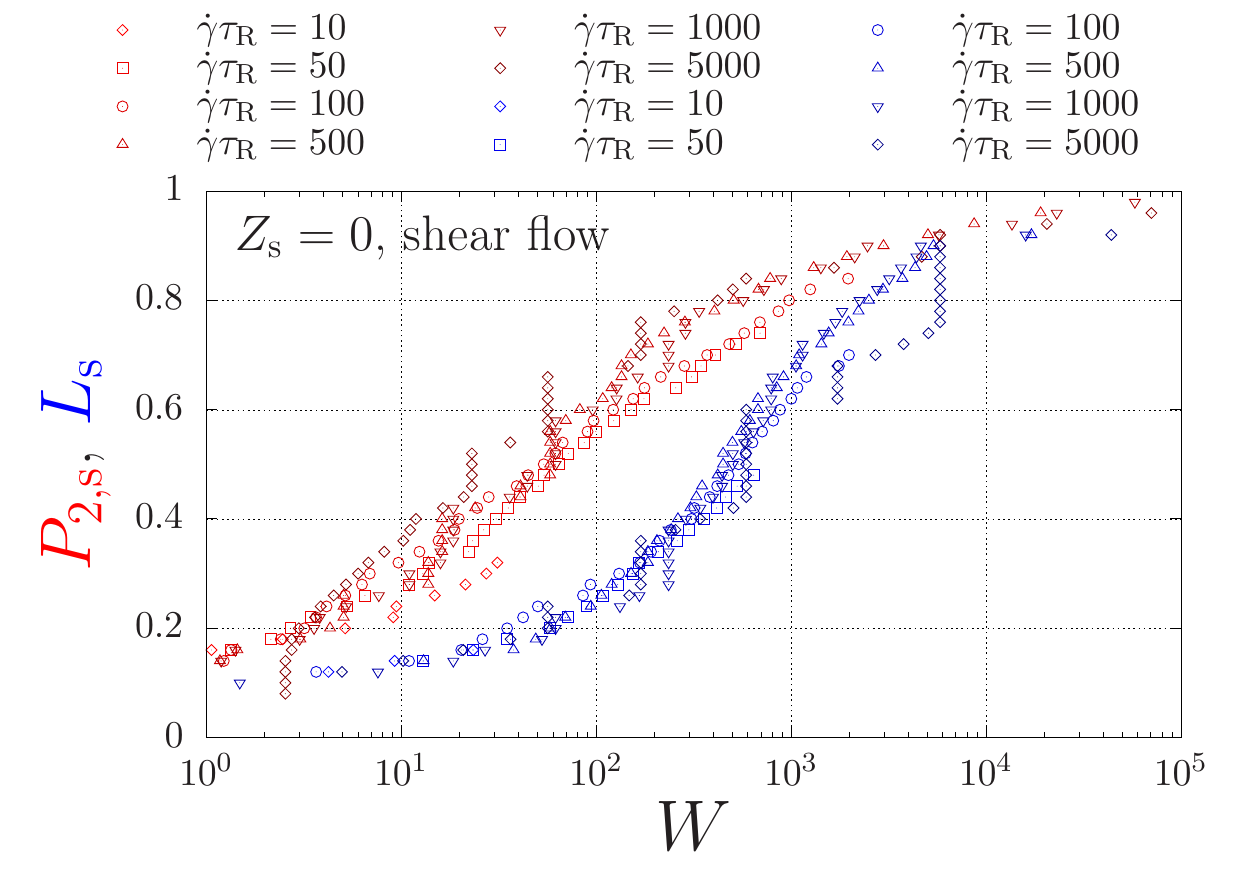}
\includegraphics*[height=5.8cm, trim={0 0 0.5cm 0}]{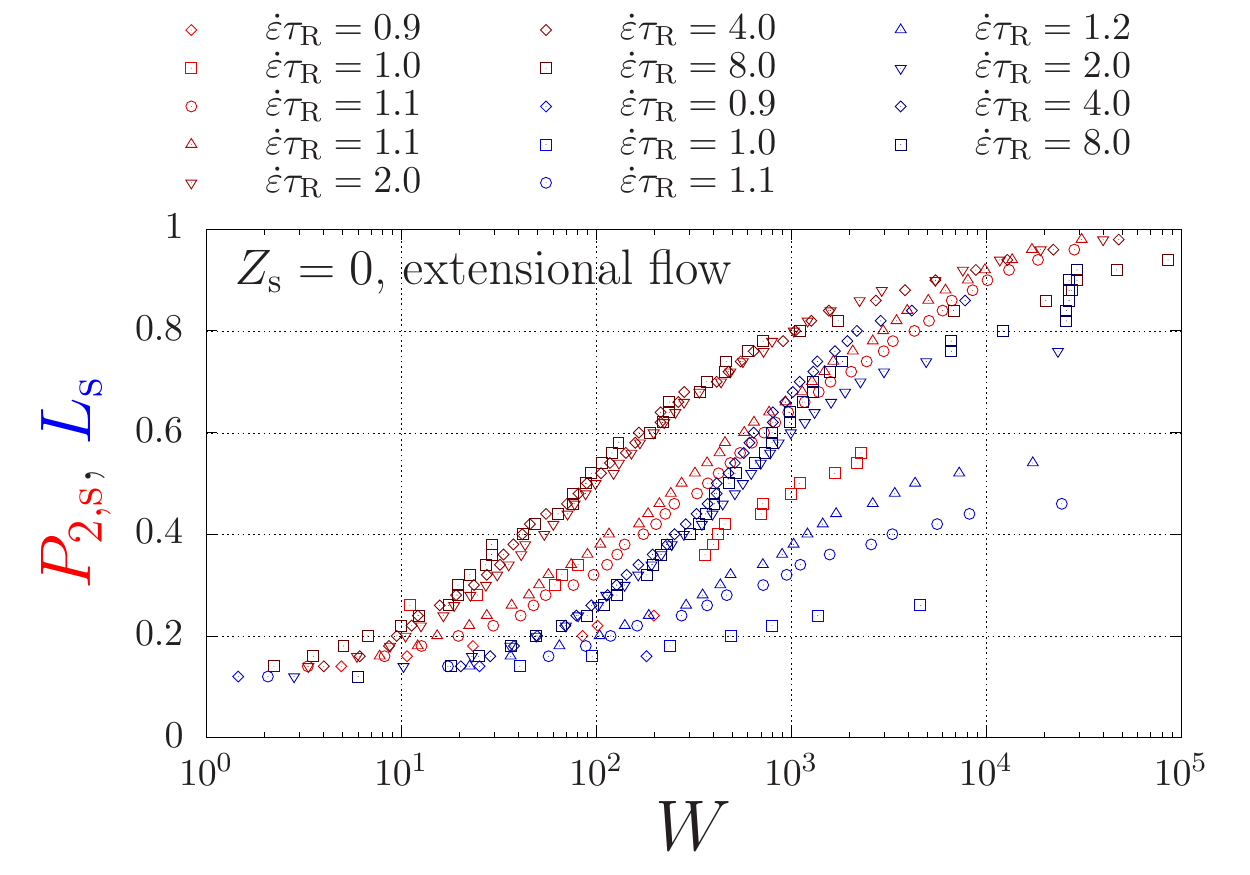}
\includegraphics*[height=5.8cm, trim={0 0 0.5cm 0}]{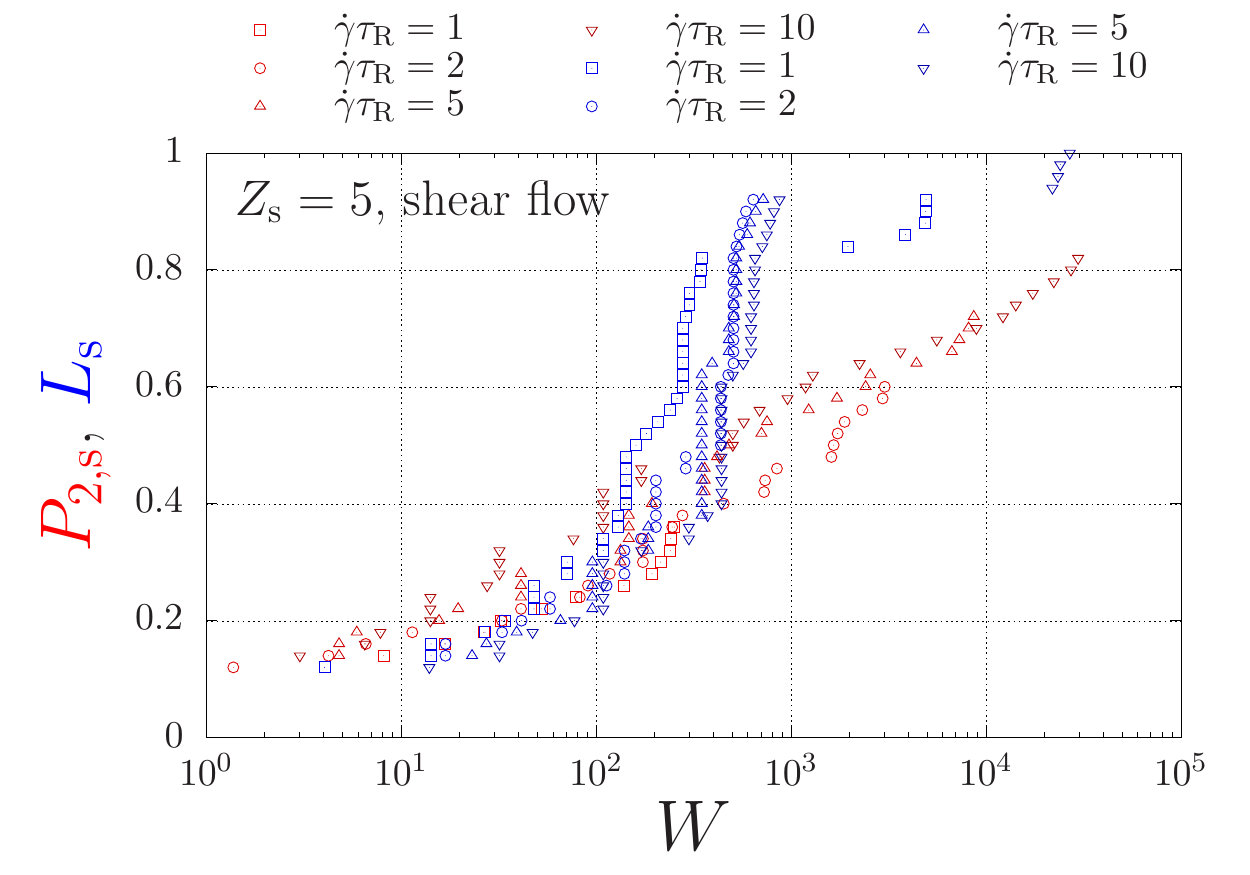}
\includegraphics*[height=5.8cm, trim={0 0 0.5cm 0}]{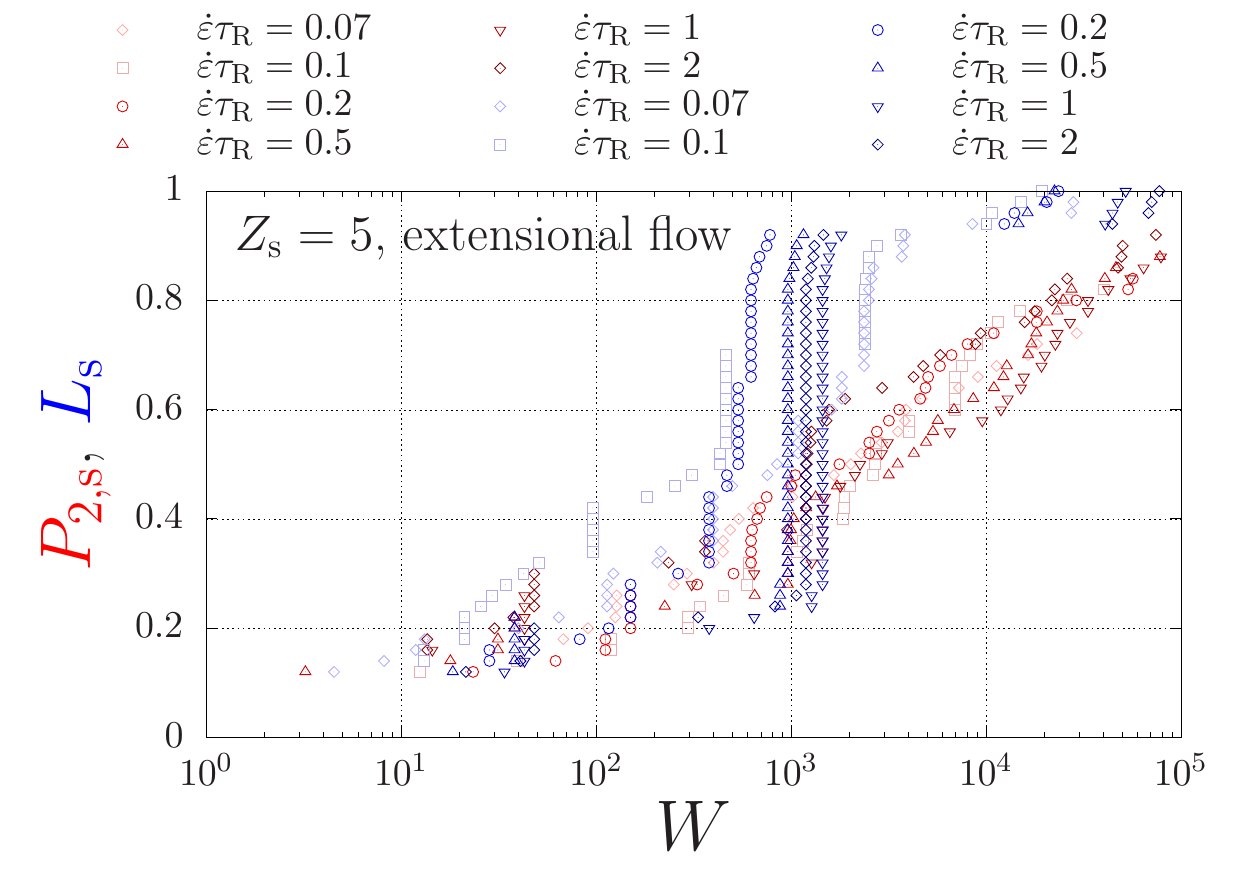}
\caption{
Nematic order parameter, $P_{2,\mathrm{s}}$ and characteristic stretch ratio, $L_\mathrm{s}$, against the specific work (see main text) for sticky (red) and non-sticky (blue) polymers in shear (left) and extensional (right) flow. The symbols are obtained from simulations with various strain rates; the curves serve as a guide to the eye. 
}
\label{fig:SpecificWork}
\end{figure}

The top panels of this Figure give the nematic order parameter, $P_{2,\mathrm{s}}$, and the measure for stretch fluctuations, $L_\mathrm{s}$ against the critical specific work.
For large values of the critical work, both measures converge, which suggests that both measures can interchangeably used as predictors for flow-induced crystallisation for non-sticky chains.
We notice that the critical work in shear (left) and extensional flow (right) show similar trends well above the stretch transition. Just above this transition the critical work required is relatively large.
This implies a monotonically decreasing critical work with an increasing strain rate, which is due to the suppression of energy dissipation by recoiling of the chains. 
This is in contrast to the typical behaviour in experiments on non-associating polymers (e.g., the flow-induced crystallisation of HDPE \cite{Holland12a}), where the critical work \emph{increases} with an increasing strain rate.
We argue this discrepancy occurs because we here consider {unentangled} rather than entangled chains.
Finally, the top panels of \Fig{fig:SpecificWork} confirm the expected behaviour where the nematic order parameter (red) is typically larger than the stretching parameter (blue): with an increasing specific work the chains first align and then stretch.

This behaviour is crucially altered for the sticky polymers, as shown in the bottom panels of \Fig{fig:SpecificWork}.
We find that the alignment of the chains requires more critical work both in shear (left) and extensional flow (right), which is due to the fact that the full alignment of the chains requires the opening of intermolecular associations.
On the other hand, the stretching of chain segments can take place before global chain alignment.
The stretching parameter (blue) follows a sharp sigmoidal dependence against the critical work, and rapidly outgrows the alignment parameter (red).  
This supports out hypothesis that flow-induced crystallisation may be achieved a small critical specific work by exploiting the stochastic nature of associating polymers.

We have investigated how the critical specific work, $W$, and the timescale, $t_\mathrm{s}$, at which the critical specific work is reached depend on the strain rate in \Fig{fig:SpecificWorkB}.
The left panel shows that the timescale scales as $t_\mathrm{s}\propto \mathrm{Wi}^{-1}$, as one may expect and discuss in more detail below.
Below the stretch transition this dependence becomes stronger: under these conditions many chain stretches are attempted, but fail due to sticker opening and lead to energy dissipation through chain retraction. 
This crossover between two regimes qualitatively agrees with that found in Figure 2 of the work by Holland et al. on silk \cite{Holland12a}; more dedicated research is needed to investigate this observation.

\begin{figure}[ht!]
\centering
\includegraphics*[height=5.8cm, trim={0 0 0.5cm 0}]{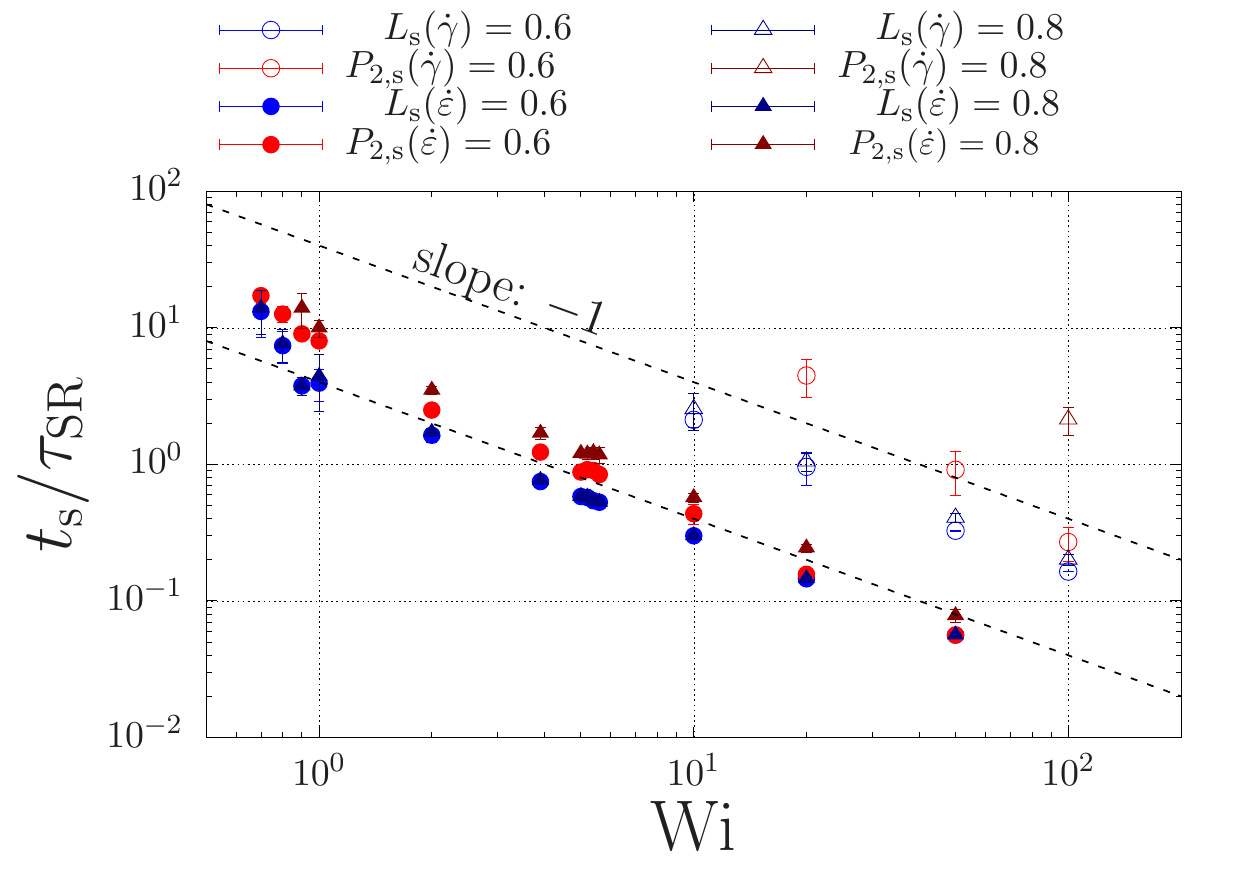}
\includegraphics*[height=5.8cm, trim={0 0 0.5cm 0}]{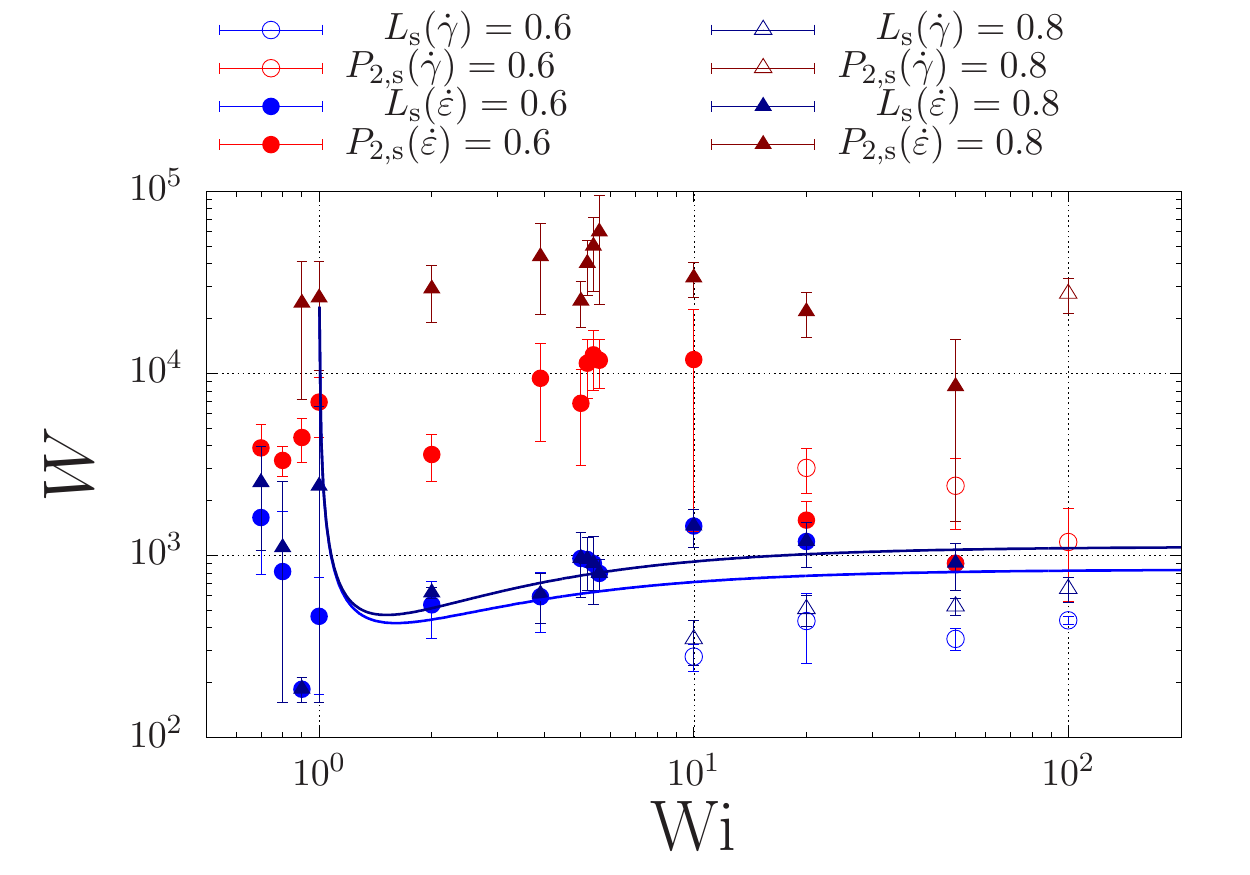}
\caption{
The critical time (left) and the specific critical work (right) against the Weissenberg number for various $L_\mathrm{s}$ and $P_{2,\mathrm{s}}$ criteria for the critical condition.
Note that $\mathrm{Wi}=1$ corresponds to the stretch transition for the sticky chains while $\mathrm{Wi}=10$ corresponds to that of the chains without stickers.
The solid lines are given by \Eq{eq:ApproxWork} for $L_\mathrm{s}=0.6$ and for $L_\mathrm{s}=0.8$.
}
\label{fig:SpecificWorkB}
\end{figure}

The right panel of \Fig{fig:SpecificWorkB} shows that the critical specific work  shifts to higher values for an increasing strain rate, which is in agreement with experimental observations on the flow-induced crystallisation of synthetic polymers and silk \cite{Holland12a}.

We explain the increase of the critical specific work  with an increasing strain rate in terms of the two-state model that we introduced in the Theory section.
We argue that the stress is dominated by the contributions of stretched chains in the closed state,
\begin{equation}
  \sigma_{xx}(t) = c\int  P_0(\lambda,t) \lambda(t)^2\mathrm{d}\lambda,
\end{equation}
with $c$ a constant, assuming that the open chains are in a relaxed state.
Here, $P_0(\lambda,t)$ is the stretch distribution of the closed chains, of which we will discuss the dynamics below.
We will then calculate the critical specific work as $W=\int_0^{t_\mathrm{s}}\sigma_{xx}\dot{\varepsilon}\mathrm{d}t$.  
To calculate $W$, we first will determine $t_\mathrm{s}$ using the criterion 
\begin{equation}
  \int_{L_\mathrm{s}\lambda_{\mathrm{max},i}}^{\lambda_{\mathrm{max},i}} P_{0}(\lambda, t_\mathrm{s}) \mathrm{d}\lambda \ge P_\mathrm{s},
\label{eq:criterion}
\end{equation}
which, as before, implies a minimum concentration of chains with a stretch ratio of at least $\lambda_\mathrm{s}=L_\mathrm{s}\lambda_{\mathrm{max},i}$.
Secondly, we will need an expression for the time evolution of the probability density $P_0$.

To obtain $P_0$, we will assume that all chains that have (temporarily) opened are sufficiently relaxed compared to the most stretched chains to have a negligible contribution to the overall stress $\sigma_{xx}$. Therefore, we will only take into account the loss of strongly stretched chains by opening rate $\kappa_-$, and ignore the contribution of closing events by rate $\kappa_+$. We will further use the initial condition $P(\lambda,0)= \delta(1-\lambda)$, with $\delta(.)$ the Dirac delta distribution. 
Using these considerations,  \Eq{eq:LaPlace_stretchingmode} of the two-state model simplifies to
\begin{equation}
  \frac{\partial \tilde{P}_0}{\partial  y}(\lambda,s)= 
- (\kappa_{-}+\dot{\varepsilon}+s)\tilde{P}_0(\lambda,s)  + c \delta(1-\lambda),
\end{equation}
of which the solution is
\begin{equation}
  P_0(\lambda,t) = \delta( \ln \lambda-\dot{\varepsilon}t )
\lambda^{-\left(1+1/{(\dot{\varepsilon}}\tau_\mathrm{s})\right)},
  \label{eq:P0approx}
\end{equation}
where we inserted $\kappa_-=\tau_\mathrm{s}^{-1}$.
This solution is a Dirac delta distribution that in time shifts to high stretch values along the $\lambda$ axis,  while its amplitude decreases due to sticker opening.
This gives rise to the simple relation between the critical stretch and the time scale by $t_\mathrm{s}=\ln\lambda_\mathrm{s}/\dot{\varepsilon}$, which is in agreement with our simulated results displayed in \Fig{fig:SpecificWork}.
We insert this equation into the expression for the critical specific work, $W=\int_0^{t_\mathrm{s}}\sigma_{xx}\dot{\varepsilon}\mathrm{d}t$, and find
\begin{equation}
W(\dot{\varepsilon})=  c \left(1-\frac{1}{\dot{\varepsilon} \tau_\mathrm{s}}\right)^{-1}
\exp\left[\left(1-\frac{1}{\dot{\varepsilon} \tau_\mathrm{s}}\right)\ln \lambda_\mathrm{s}-1\right]
,\quad\text{for}\quad\dot{\varepsilon}>\dot{\varepsilon}_\mathrm{min},\label{eq:ApproxWork}
\end{equation}
where $\dot{\varepsilon}_\mathrm{min}$ is the minimum strain rate for which the criterion in \Eq{eq:criterion} is obeyed.
{
This function diverges at $\dot{\varepsilon}\tau_\mathrm{s}=1$, but with an increasing strain rate decreases to a minimum, following which it monotonically increases towards a plateau value.}
Physically, this plateau value represents the case where the entire distribution of chains is stretched to reach the critical stretch value $\lambda_\mathrm{s}$.
In this case, the concentration of stretched segments far exceeds the critical concentration, and more energy has been put into the system then needed.
By decreasing the strain rate, an increasing number of stickers are able to open and the stress is relaxed, in turn decreasing the critical specific work to achieve the critical condition in \Eq{eq:criterion}.
This supports our proposition that the stochastic nature of the binding and unbinding of associations enables to 
molecularly engineer associating polymers to undergo flow-induced crystallisation at low energetic costs.
{In particular, we have shown, using simulations and an approximate theory in \Eq{eq:ApproxWork}  that there is an \emph{optimum} strain rate at which the critical work for critical stretch is minimised}

\section{Conclusions}\label{sec:Conclusions}

This work has shown that the transient evolution of the chain-stretch distribution of associating 'sticky' polymers in shear, and especially extensional, flow possesses an extremely rich structure. The theoretical and numerical investigations reported here were  driven by the observation that the silk protein (i) undergoes efficient, chemically tunable, flow-induced crystallisation and (ii) can be modelled as an associating/sticky polymer.
Our findings have implications for the interpretation of silk-spinning data, as well as to the development of novel associating polymers and the computational modelling tools (we introduced  a `sticky' sliplink model, and an analytical two-state master equation which may be transferable to also address the peculiar dynamics of ring polymer in flow \cite{ HuangQ19, OConnor20, OConnor21, Bonato21}).

Regarding silk rheology, we have confirmed our hypothesis that the stickers between chains may reduce the critical specific work to induce flow-induced crystallisation under reasonable assumptions for critical crystallisation criteria.
While intuitively this is achieved by the presence of stickers shifting the stretch transition to smaller strain rates (which is of course highly relevant to the processing conditions both in animals and in the industry), the actual reason for creating a minimum in critical work is more subtle:
The stickers fix a network at time scales below the sticker dissociation time, and in fact hinder the alignment of the chains in the direction of the flow field.
We found, using the  measure $L_\mathrm{s}$, that owing to the stochastic nature of the binding and unbinding of the stickers a significant portion of the chain segments can be stretched to a large extend, while the global alignment, measured by the  nematic order parameter $P_{2,\mathrm{s}}$, remains small.
Assuming that crystallisation commences above a critical concentration of stretched chain segments, rather than when the entire population is stretched, we found a significant reduction of the critical specific work is possible.

Focussing on our finding that chain alignment at low, non-stretching, flow rates requires less specific work in the absence of stickers (and presumably for low sticker lifetimes) than with stickers, while the stretching of the chains at high rates is helped by long sticker lifetimes, we argue that control over both the structural aspects of the final material and over the specific work needed is possible through time- or position-dependent sticker lifetimes.
We argue this can be achieved through external chemical control.
Indeed, during its larval life cycle, the silkworm stably stores its silk solution at a high viscosity, but just prior to silk spinning it lowers the viscosity through an increase of the potassium concentration through a decreasing lifetime of calcium bridges (stickers) \cite{Laity18,Schaefer20}.
This, as we can now interpret as a mechanism to ease chain alignment in flow.
Intriguingly, downstream the spinning duct the acidity increases \cite{Koeppel21}, which we expect to increase the stability and hence the lifetime of the calcium bridges, and hence enhance local chain stretching, see \Fig{fig:Hypothesis}, which may in turn disrupts the solvation layer of the protein and induce efficient crystallisation \cite{Holland12a, Dunderdale20}.

While this seems a compelling mechanism for efficient flow-induced crystallisation, it is not yet clear how this process may be optimised.
In the case of \emph{Bombyx mori} silk, we identified a regular spacing of the negative charges along the backbone of the chain, with strands of approximately $500$ uncharged amino acids between; the length of these sticker strands is of the order of the entanglement molecular weight.
The regularity of the spacing and the coincidental similarity between the number of stickers and entanglements suggests some degree of evolutionary optimisation.
The functionality of ordered- versus random co-polymers is of high importance from a synthetic polymer chemistry point of view, and needs to be addressed using simulations that include both associations and entanglements.

{We conclude that our modelling approach leaves us well prepared to investigate the ways in which the evolution of silk-producing organisms may have exploited the potential optimal strategies for efficient fibre processing.
The next piece of physics to add to this account of the rheology of polymers with temporary assocations, not only for modelling silk proteins but also general associating polymers, concerns the interaction between entanglements and associations in strong flow.
We anticipate that this will further enrich the ongoing debate in polymer physics on the physics of entanglement generation and destruction (i.e., `entanglement stripping') in non-linear rheo-physics, as well as continue the account of how silk-forming organisms point to novel rheo-physics of flow-induced phase-transformations.}

\section{Appendix}

\subsection{Algorithm}

Because of the large distribution of chain stretch in the conditions we are interested in, there is also a large distribution of opening rates; in our previous work we used small time steps in which the chain conformation was updated, and each closed pair had a sufficiently small opening probability.
Here, we significantly improve this algorithm by enabling much larger time steps between conformational updates, and during which the stickers may open and close many times, see \Fig{fig:Algorithm}.

\begin{figure}[ht!]
\centering
\includegraphics*[width=12cm]{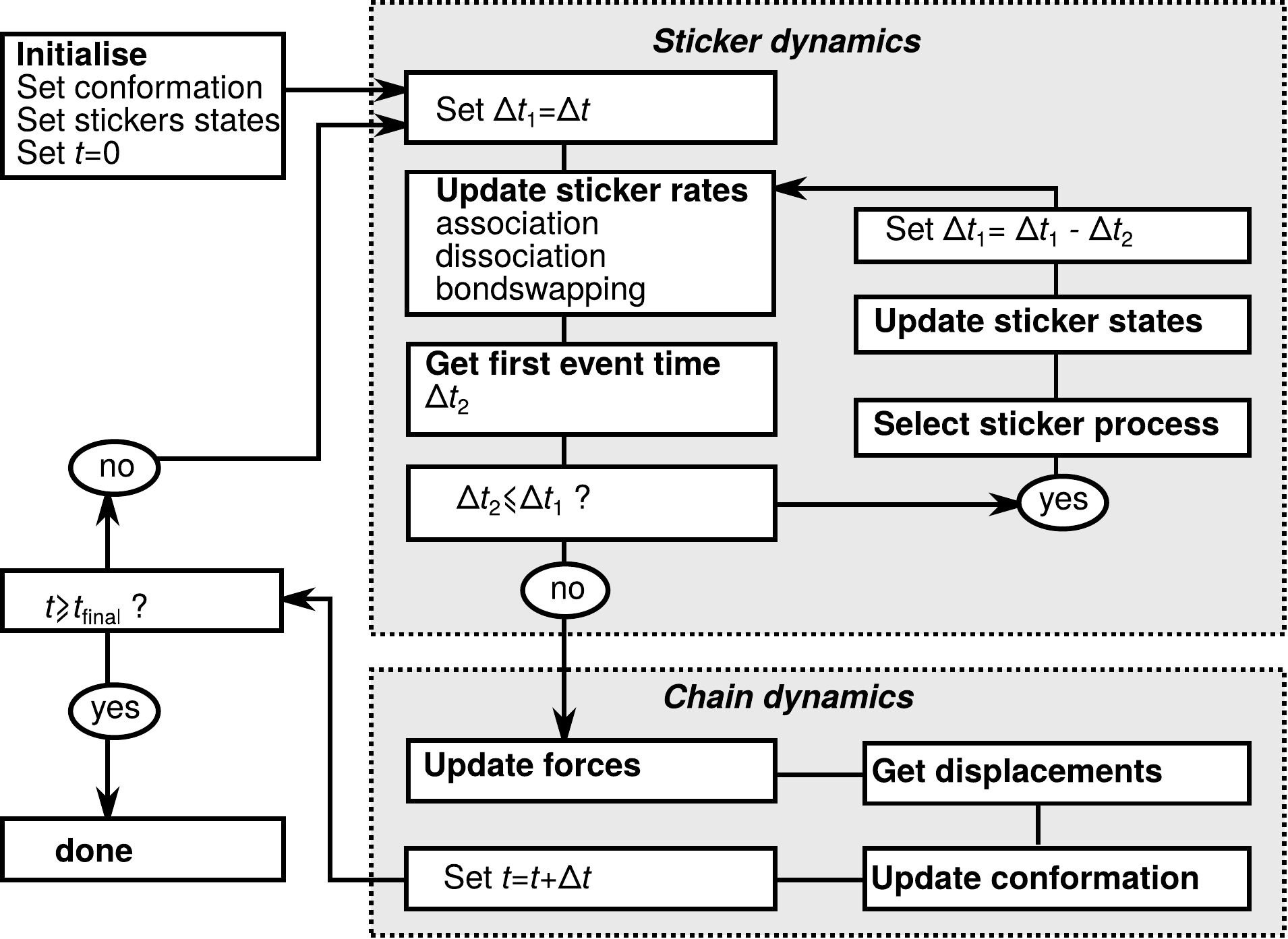}
\caption{Flow chart of the algorithm so simulate the conformational dynamics of sticky polymers and the dynamics of sticker association and dissociation.}
\label{fig:Algorithm}
\end{figure}

In our algorithm, we update the \emph{chain conformation} using the Brownian dynamics equation from the previous section using a time span $\Delta t$.
Depending on the opening and closing rates, during this time span, $\Delta t_1\equiv \Delta t$, the \emph{sticker configuration} may be updated many times or not at all according to a kinetic Monte Carlo (kMC) scheme \cite{Lukkien98, HeermannBook10, JansenBook12}.
In every kMC step, the rate at which any opening or closing event may occur is calculated as  $W_\mathrm{T}=W_\mathrm{a}+W_\mathrm{d}$, with
\begin{equation}
  W_\mathrm{a} = k_\mathrm{a}N_\mathrm{open}(N_\mathrm{open}-1)/2,
\end{equation}
the sum of closing rates and 
\begin{equation}
  W_\mathrm{d} = \sum_{q=1}^{N_\mathrm{closed}/2}k_\mathrm{d,q},
\end{equation}
the sum of dissociation rates, where $k_\mathrm{d,q}$ differs for the different sticker pairs due to dispersity in chain tension.
In these expressions, $N_\mathrm{open}$ and $N_\mathrm{closed}$ are the number of open and closed stickers, respectively; $N_\mathrm{open}(N_\mathrm{open}-1)/2$ is the total number of possible associations, and the index $q$ sums over all $N_\mathrm{closed}/2$ \emph{pairs} of closed stickers.
Using this sum of rates, the time $\Delta t_2$ at which the first opening or closing event  occurs is
\begin{equation}
  \Delta t_2 = -\frac{1}{W_\mathrm{T}}\ln(u),
\end{equation}
with $u\in(0,\,1]$ a uniform random number (our code uses random numbers using the open-source SFMT library \cite{Saito08}).
If $\Delta t_2$ exceeds the time span $\Delta t_1$, no opening or closing events occurs. 
However, if $\Delta t_2<\Delta t_1$ then a second random number $\in[0,\,1]$ is drawn, and a closing event is selected with probability $k_\mathrm{a}/W_\mathrm{T}$, and a dissociation event $q$ is selected with probability  $k_\mathrm{d,q}/W_\mathrm{T}$.
After updating the configurations of the stickers, the time span is updated to $\Delta t_1=\Delta t_1+\Delta t_2$. 
The kMC scheme is terminated when  $\Delta t_2>\Delta t_1$, following which the chain conformation is updated.

While in the linear rheological conditions we solve the dynamics using a fixed time step, in strong flow we implemented an adaptive time step to handle the large and fast fluctuations in stretch that emerge in some parameter regimes of the system.
In every iteration $n$, the time step for the next iteration is updated as
\begin{equation}
  \Delta t^{n+1} = \Delta t^{n}\left(\min_{Q_i}\frac{\mathrm{tolerance}}{\mathrm{error}}\right)^{0.25},
\end{equation}
where an error and tolerance are calculated for the change of each end-to-end vector $\mathbf{Q}_i$.
We defined the error value for each change in $Q_i$ as $\mathrm{error}=|\Delta Q_i^{n}|/Q_\mathrm{max}$, with $Q_\mathrm{max}$ set by $\lambda_\mathrm{max}$.
For the tolerance value we use scalar values $\mathrm{tol}_{-}$ and $\mathrm{tol}_{+}$ depending on whether $|Q_i^{n}|$ is smaller or larger than a certain cutoff set by $\lambda_\mathrm{cutoff}<\lambda_\mathrm{max}$.
Above the cutoff, we avoid numerical instabilities due to the singularity at  $\lambda_\mathrm{max}$ by using
\begin{equation}
  k_\mathrm{s}(\lambda>\lambda_\mathrm{cutoff}) =
k_\mathrm{s}(\lambda_\mathrm{cutoff})\times \left(\frac{\lambda}{\lambda_\mathrm{cutoff}}\right)^\alpha.
\end{equation}
For continuity of the derivative, $\alpha=4 c^2/(3-4c^2+c^4)$, with $c=\lambda_\mathrm{cutoff}/\lambda_\mathrm{max}$; for a cutoff $\lambda_\mathrm{cutoff}=0.9\lambda_\mathrm{max}$ even this smooth potential is steep ($\alpha \approx 8$), and in practice we use a softer potential ($\alpha=4$).

\subsection{Long-time limit of the two-state model}

To solve Eqs.~(\ref{eq:TwoStateEqA}-\ref{eq:TwoStateEqB}) in the long-time limit, we make the useful approximation that at an intermediate time $t=t_\ast$ the distribution is equilibrated for small stretches $\lambda\leq\lambda_\ast$, while the large-stretch tail of the distribution is unoccupied.
Hence, at $t=t_\ast$ the distribution is given by
\begin{align}
  P_{0}(0,y)&=\frac{c'}{c}  P_0^{\mathrm{eq}} \Theta(-y+y_\ast)\\
  P_{1}(0,y)&=\frac{c'}{c}  P_1^{\mathrm{eq}} \Theta(-y+y_\ast),
\end{align}
where $y_\ast \equiv \ln \lambda_\ast$ and where $\Theta$ is the Heaviside step function.
The prefactor
\begin{equation}
  c'= \left(  1+c\frac{1}{1+\nu}\mathrm{e}^{(1+\nu)y_\ast}  \right)^{-1}>1
\end{equation}
normalises the distribution.
We now set $\lambda_\ast$ to a large value, so $c'\approx c$, and at late times $t>t\ast$ the filling of the tail of the distribution (for $\lambda>\lambda_\ast$) occurs with a negligible effect on the distribution at small stretches.

Using these approximations, we can now solve the dynamic equations for $\lambda>\lambda_\ast\gg 1$ using the Laplace transform $\tilde {P_i}(s,y) \equiv \mathcal{L}\left\{{P}_i(t,y)\right\}$ for $i=0,1$.
Using the standard Laplace transform of the time derivative $\mathcal{L}\left\{\partial {P}_i/\partial t\right\}=s\tilde{P}_{i}(s,y) - P_i(0,y)$, we get
\begin{alignat}{3}
  \frac{\partial \tilde{P}_0}{\partial  y} &= 
- (\kappa_{-}+\dot{\varepsilon}+s)\tilde{P}_0 
&+ \kappa_{+}\tilde{P}_1 &+ P_0(0,y)/s,\label{eq:LaPlace_stretchingmode}\\
  \frac{\partial \tilde{P}_1}{\partial  y} &=  +  \kappa_{-} \tilde{P}_0 
&- (\kappa_{+}+\dot{\varepsilon}+s-\tau_\mathrm{R}^{-1})\tilde{P}_1 
&+ P_1(0,y)/s,
\end{alignat}
of which the solution is of the form
\begin{align}
  \tilde{P}_0(s,\lambda) &= c_0^{+}(s) \lambda^{\nu_+(s)} + c_0^{-}(s) \lambda^{\nu_-(s)}\\
  \tilde{P}_1(s,\lambda) &= c_1^{+}(s) \lambda^{\nu_+(s)} + c_1^{-}(s) \lambda^{\nu_-(s)},
\end{align}
with $\nu_-(s)$ and $\nu_+(s)$ the eigenvalues given by
\begin{align}
  \nu_{\pm}=
\frac{1}{2\dot{\varepsilon}(1-\dot{\varepsilon}\tau_\mathrm{R})}
&\Bigl(
(2\dot{\varepsilon}+\kappa_-)(1-\dot{\varepsilon}\tau_\mathrm{R})-\dot{\varepsilon}\tau_\mathrm{R}\kappa_+
+s(1-2\dot{\varepsilon}\tau_\mathrm{R})
\Bigr.\nonumber\\
&\Bigl.
\pm \sqrt{
(s+\kappa_-(1-\dot{\varepsilon}\tau_\mathrm{R}))^2
+2\dot{\varepsilon}\tau_\mathrm{R}(s-(1-\dot{\varepsilon}\tau_\mathrm{R})\kappa_-)\kappa_++(\dot{\varepsilon}\tau_\mathrm{R}\kappa_+)^2
}
\Bigr),\label{eq:Eigenvalue}
\end{align}
and where the coefficients, $c_i^{\pm}$, follow from the boundary condition at $y=y_\ast$.

At late times, i.e., for small $s$, we have  $\nu_-(s)\approx  \nu_\mathrm{eq} - (s/\dot{\varepsilon})\nu'+(1/2)(s/\dot{\varepsilon})^2\nu''$, where $\nu_\mathrm{eq}$ is given by \Eq{eq:StretchingExponent}, 
{and where
\begin{equation}
  \nu' \equiv \left.\frac{\mathrm{d}\nu}{\mathrm{d}(s/\dot{\varepsilon})}\right|_{s=0} =\left(1 -  \frac{1}{1-\mathrm{Wi}^0} 
\frac{1}{1 - \mathrm{Wi}^{\mathrm{sticky}}}\right), 
\end{equation}
and
\begin{equation}
  \nu'' \equiv 2p\mathrm{Wi^0}\frac{
         \mathrm{Wi}^{\mathrm{sticky}}}{
        (1-\mathrm{Wi}^{\mathrm{sticky})})^3},
\end{equation}
are both positive, provided that the sticky Weissenberg number is sufficiently small, $\mathrm{Wi}^{\mathrm{sticky}}\equiv \mathrm{Wi}^0/(1-p)<1$ \cite{Schaefer21A}, where $\mathrm{Wi}^0=\dot{\varepsilon}\tau_\mathrm{R}$ is the Weissenberg number of the chain without stickers.}

From the boundary condition, we find that the coefficients must be of the form $c_i^{\pm}\propto 1/s$.
As the `+' solution leads to a non-normalisable solution, however, $c_i^{+}=0$, and the solution is 
\begin{align}
  \tilde{P}_0(s,\lambda) &=  \frac{c}{s} (\lambda/\lambda_\ast)^{
\nu_--(s/\dot{\varepsilon})\nu'-\frac{1}{2}(s/\dot{\varepsilon})^2\nu''+\mathcal{O}(s^3)},\label{eq:LaplaceSolution}\\
  \tilde{P}_1(s,\lambda) &=  \frac{\kappa_{+}}{\kappa_{-}}\frac{\dot{\varepsilon}}{(\dot{\varepsilon}-\tau_\mathrm{R}^{-1})} \tilde{P}_0(s,\lambda).
\end{align}
Finally, after taking the inverse Laplace transform, we have 
\begin{align}
  {P}_0(t,\lambda) &=  c \left(\frac{\lambda}{\lambda_\ast(0)}\right)^{\nu_\mathrm{eq}}\Theta(\nu'\ln\lambda/\lambda_\ast - \dot{\varepsilon}t)\\
  {P}_1(t,\lambda) &=  \frac{\kappa_{+}}{\kappa_{-}}\frac{\dot{\varepsilon}}{(\dot{\varepsilon}-\tau_\mathrm{R}^{-1})} {P}_0(t,\lambda).
\end{align}
Hence, the exponentially extending front of the distribution is located at the stretch ratio
\begin{equation}
  \lambda_\ast(t)=\lambda_\ast(0)\exp\left[\left(1 -  \frac{1}{1-\mathrm{Wi}^0} 
+ \frac{1}{1 - \mathrm{Wi}^{\mathrm{sticky}}}\right)^{-1}\dot{\varepsilon}(t-t_\ast)\right].
\end{equation}

{We have checked the validity of our interpretation of a narrow moving-front by also calculating the width of this front.
To do this, we consider the relaxation function $f(t)=P(y,t)/P_{\mathrm{eq}}(y)$ with again $y=\ln\lambda$, and $P$ and $P_\mathrm{eq}$ the transient and steady-state stretch distributions, respectively.
A narrow front that reaches $y$ at time $\tau$ and equilibrates at time $\tau+\Delta$ may be approximated by 
\begin{equation}
f(t)=\begin{cases}
  0,\quad\text{for}\quad t<\tau\\
  (t-\tau)/\Delta,\quad\text{for}\quad \tau \leq t<\tau+\Delta\\
  1,\quad\text{for},\quad t\geq\tau+\Delta.
\end{cases}
\end{equation}
The Laplace transform of this function is
\begin{equation}
  \mathcal{L}\{f\}=\frac{1}{s^2\Delta}\mathrm{e}^{-s\tau}\left(1-\mathrm{e}^{-s\Delta}\right).
\end{equation}
We compare this result to the solution of the two-state model in \Eq{eq:Eigenvalue} through a second-order Taylor expansion of the exponential terms
\begin{equation}
  \mathcal{L}\{f\}=\frac{1}{s}\Bigl( 1-\underbrace{(\tau+\frac{1}{2}\Delta)}_{(\nu'/\dot{\varepsilon})\ln y}s+\frac{1}{2}\underbrace{(\tau^2+\frac{1}{3}\Delta^2+\Delta\tau)}_{(\nu''/\dot{\varepsilon}^2)\ln y}s^2\Bigr).
\end{equation}
From the linear term, we find $\tau+\Delta /2 = (\nu'/\dot{\varepsilon})\ln y$ (as expected from \Eq{eq:MovingFront}).
After substitution into the second term and elimination of this variable, we find the width of the front to be 
\begin{equation}
  \Delta = \sqrt{12}\sqrt{(\nu''/{\dot{\varepsilon}})\ln y-(\nu'/{\dot{\varepsilon}})^2\ln y}.
\end{equation}
The relative width, compared to the location of the front ($\tau+\Delta /2$), is
\begin{equation}
  \Delta_\mathrm{rel}\equiv\frac{\Delta}{\tau+\Delta/s} = \sqrt{12}\sqrt{\frac{\nu''}{(\nu')^2}-1}\label{eq:RelativeWidth}
\end{equation}
The relative width calculated in the time-domain also represents the relative width of the (logarithmic) stretch distribution:
\begin{equation}
  \Delta_\mathrm{rel}\equiv\frac{y(\tau+\Delta)-y(\tau)}{y(\tau+\Delta/2)}.
\end{equation}
Upon approaching the strain rate where the mean stretch diverges, i.e., at $\mathrm{Wi}^{\mathrm{sticky}}=1$, the relative width of the  front diverges as $\Delta_\mathrm{rel} \approx \sqrt{24p \mathrm{Wi}^0\mathrm{Wi}^{\mathrm{sticky}}/(1-\mathrm{Wi}^{\mathrm{sticky}})}$.
In this equation, the bare Weissenberg number is $\mathrm{Wi}^0=\mathrm{sticky}(1-p)\tau_\mathrm{R}/\tau_\mathrm{s}$. Hence, if the sticker lifetime is $\tau_\mathrm{s}=10\tau_\mathrm{R}$ and the fraction of closed stickers is $p=0.9$ (as in our simulations), then significant broadening of the front only happens very close to the stretch transition: $\mathrm{Wi}^{\mathrm{sticky}}>0.99$.
This verifies that our approximation of a narrow front is indeed accurate.} 

\subsection{Power-law regression}

To determine the sticky Rouse diffusivity, $D_\mathrm{SR}$, from the mean-square displacement of the centre of mass 
\begin{equation}
  \ln \mathrm{MSD}=\ln(6D_\mathrm{SR}) + \ln t
\end{equation}
as a function of time $t$, and the stretch exponent, $\nu$, from the probability distribution
\begin{equation}
  \ln P= c  + \nu \ln \lambda
\end{equation}
as a function of the stretch ratio, $\lambda$, we write both equations in the form 
\begin{equation}
  y = a + b x
\end{equation}
and perform common linear regression. However, because both power-laws represent asymptotic behaviour for large $x$, there is also a cutoff value, $x_\mathrm{cutoff}$, above which they apply.
We determine the cutoff by minimising
 \begin{equation}
  \chi^2(a, b, i_0) \equiv \frac{1}{N_\mathrm{data}+1-i_0-N_\mathrm{par}} 
  \sum_{i=i_0}^{N_\mathrm{data}}\frac{
(y_i^{\mathrm{data}} - y_i^{\mathrm{fit}}(a,b) )^2}{\sigma_i^2},
\end{equation}
with respect to $a$, $b$ and $i_0$ (note that $x_{i_0} = x_\mathrm{cutoff}$); $\sigma_i$ is the uncertainty on the simulated $y$ data.
Here, we set $b=1$ fixed and the number of free parameters $N_\mathrm{par}=1$ for extracting the sticky Rouse diffusivity from the MSD data.
To determine the stretch exponent ($\nu$) from the stretch distributions we use the same approach, but leave $b$ as a free fitting parameter and set $N_\mathrm{par}=2$.

\begin{acknowledgments} 
This research was funded by the Engineering and Physical Sciences Research Council [grant number EPSRC (EP/N031431/1)]. C.S. is grateful for the computational support from the University of York high-performance computing service (the Viking Cluster), and thanks Daniel J. Read for his suggestion to pair closed stickers of different chains. C.S. and T.C.B.M. thank Pete Laity and Chris Holland for useful discussions on the mechanism of silk self-assembly. 
\end{acknowledgments}   

\bibliography{references} 
\end{document}